\documentclass{aa}

\usepackage{graphicx}
\usepackage{txfonts}
\usepackage{url}
\usepackage{natbib}
\usepackage{color}
\bibpunct{(}{)}{;}{a}{}{,}
\newcounter{notetab}
\newcommand{\zero}{\setcounter{notetab}{0}}

\begin{document}
\title{Variation in dust properties in a dense filament of the Taurus molecular complex (L1506)}

\author{N. Ysard\inst{\ref{inst1}}
\and A. Abergel\inst{\ref{inst1}}
\and I. Ristorcelli\inst{\ref{inst2}}
\and M. Juvela\inst{\ref{inst3}}
\and L. Pagani\inst{\ref{inst4}}
\and V. K\"onyves\inst{\ref{inst1}, \ref{inst5}}
\and L. Spencer\inst{\ref{inst6}}
\and G. White\inst{\ref{inst7}, \ref{inst8}}
\and A. Zavagno\inst{\ref{inst9}}}
\institute{IAS, CNRS (UMR8617), Universit\'e Paris Sud, B\^at. 121, F-91400 Orsay, France, \label{inst1}\email{nathalie.ysard@ias.u-psud.fr}
\and IRAP, CNRS (UMR5277), Universit\'e Paul Sabatier, 9 avenue du Colonel Roche, BP 44346, F-31028 Toulouse cedex 4, France\label{inst2}
\and Department of Physics, P.O. Box 64, FI-00014, University of Helsinki, Finland\label{inst3}
\and LERMA, CNRS (UMR8112), Observatoire de Paris, 61 avenue de l'Observatoire, F-75014 Paris, France\label{inst4}
\and Laboratoire AIM, CEA/DSM-CNRS-Universit\'e Paris Diderot, IRFU/Service d'Astrophysique, CEA Saclay, Orme des Meurisiers, 91191 Gif-sur-Yvette, France\label{inst5}
\and School of Physics and Astronomy, Cardiff University, Queens Buildings, The Parade, Cardiff CF24 3AA, United Kingdom\label{inst6}
\and The Open University, Department of Physics and Astronomy, Milton Keynes MK7 6AA, United Kingdom\label{inst7}
\and The Rutherford Appleton Laboratory, Chilton, Didcot, Oxfordshire OX11 0NL, United Kingdom\label{inst8}
\and Laboratoire d'Astrophysique de Marseille, CNRS/INSU Universit\'e de Provence, 13388 Marseille cedex 13, France\label{inst9}}

\abstract{}
{We observed the L1506 filament, which is located in the Taurus molecular complex, with the Herschel PACS and SPIRE instruments. Our aim is to prove the variation in grain properties along the entire length of the filament. In particular, we want to determine above which gas density this variation arises and what changes in the grain optical properties/size distribution are required.}
{We use the 3D radiative transfer code CRT, coupled to the dust emission and extinction code DustEM, to model the emission and extinction of the dense filament. We test a range of optical properties and size distributions for the grains: dust of the diffuse interstellar medium (interstellar PAHs and amorphous carbons and silicates) and both compact and fluffy aggregates.}
{We find that the grain opacity has to increase across the filament to fit simultaneously the near-IR extinction and Herschel emission profiles of L1506. We interpret this change to be a consequence of the coagulation of dust grains to form fluffy aggregates. Grains similar to those in the diffuse medium have to be present in the outer layers of the cloud, whereas aggregates must prevail above gas densities of a few 10$^3$~H/cm$^3$. This corresponds to line-of-sights with visual extinction in the V band of the order of 2 to 3. The dust opacity at 250~$\mu$m is raised by a factor of 1.8 to 2.2, while the grain average size is increased by a factor of 5. These exact numbers depend naturally on the dust model chosen to fit the data. Our findings agree with the constraints given by the study of the gas molecular lines. Using a simple approach, we show that the aggregates may have time to form inside the filament within the cloud lifetime. Our model also characterises the density structure of the filament, showing that the filament width is not constant along L1506 but instead varies by a factor of the order of 4.}
{We confirm the need for an increase in the far-IR dust opacity to explain the emission and extinction in L1506C, which we interpret as being due to dust growth. We also show that this opacity variation is valid along the entire length of the L1506 dense filament.}

   \authorrunning{N. Ysard et al.}
\titlerunning{Evolution in dust properties in a dense filament of the Taurus molecular complex (L1506)}

\maketitle

\section{Introduction}
\label{section_introduction}

Recent observations have shown that prestellar cores preferentially form inside the dense molecular filaments observed both with molecular gas lines and the thermal dust far-IR/submm emission \citep{Hartmann2002, Andre2010, Menshchikov2010, Hill2011, Nguyen2011}. It is thus important to characterise the physical properties of dust and gas in the filamentary structures of the interstellar medium (ISM) and to investigate how the star-formation process depends on these initial conditions. Only with knowledge of these properties will we be able to derive reliable and quantitative information about the dense filaments such as the way they are formed, their masses, their structures, or their evolutionary stages (disrupting, collapsing, or fragmenting into cores). The dust grains are a key factor for the ISM evolution as they control, for example, the heating of the gas and the formation of H$_2$ in the diffuse and dense ISM, respectively.

The dust emission in filaments has been thoroughly observed in a wide wavelength range during the past twenty years resulting in a few well-established facts. Firstly, it has been shown that the dust temperature decreases from the diffuse ISM ($n_H \lesssim 100$~H/cm$^3$) towards the centre of the dense molecular clouds, where $n_H \gtrsim 10^4-10^5$~H/cm$^3$, \citep{Laureijs1991, Bernard1999} and that this decrease cannot be explained only by the extinction of the radiation field. Secondly, this decrease most often comes with a decrease in the mid- to far-infrared (IR) dust emission ratio \citep{Abergel1994, Abergel1996, Bernard1999}. Another striking fact is the joint increase in the observed dust opacity with density, which is preferentially observed at far-IR and submillimetre wavelengths \citep{Cambresy2001, Stepnik2003, Kramer2003, Flagey2009}. This increase has been confirmed by a number of recent observations made with Planck, BLAST (Balloon Borne Large Aperture Submillimetre Telescope), and Herschel \citep{JuvelaCC2011, PlanckAbergel2011, Martin2012, Fischera2012b, Roy2013}. They proved an increase of at least a factor of 2 at 250~$\mu$m from the diffuse ISM to the centre of the molecular clouds, assuming that the dust emission is well approximated by a single modified blackbody. The study of \citet{Ysard2012} showed that such an increase cannot be due to radiative transfer effects but should originate in intrinsic variations in the dust properties. These aforementioned observational facts are usually attributed to grain growth occurring as a consequence of coagulation, which entails the disappearance of the smallest grains that emit in the mid-IR \citep{Ossenkopf1994, Stognienko1995, Koehler2011, Koehler2012}. Coagulation produces larger grains on average with higher absorption cross-sections at long wavelengths, which explains both the lower temperatures and the increased opacity. If they manage to become large enough ($a \sim 0.5-1$~$\mu$m), these grains are also expected to scatter light efficiently in the mid-IR (i.e., coreshine). This has recently been observed first in L183 and subsequently towards many molecular clouds \citep{Pagani_core2010, Steinacker2010, Stutz2010}.

In this paper, we study a cloud in the Taurus molecular complex, L1506, which has been observed with the ESA Herschel Space Observatory \citep{Pilbratt2010}. The dust emission at one location across this filament (L1506C, $\alpha_{2000} = 4$h18m50s, $\delta_{2000} = +25$\degr19\arcmin43.6\arcsec) was previously modelled in detail by \citet{Stepnik2003} using IRAS (InfraRed Astronomical Satellite) and PRONAOS balloon-borne experiment observations (PROgramme NAtional d'AstrOnomie Submillim\'etrique). This analysis provided direct observational evidence for the first time of a significant increase in the dust opacity in the densest part of the cloud compared to the diffuse surrounding medium. They concluded that the dust opacity had to be multiplied by a factor of $3.4^{+0.3}_{-0.7}$ above a density threshold of $n_H = (3 \pm 1) \times 10^3$~cm$^{-3}$ (corresponding to $A_{{\rm V}} = 2.1 \pm 0.5^{\rm m}$). This was interpreted as a result of the formation of fluffy dust aggregates via coagulation processes, as predicted by dust model calculations \citep{Ossenkopf1994, Ossenkopf1993, Bazell1990, Wright1987}. However, an increase of 3.4 seems difficult to explain from a theoretical point of view. For instance, \citet{Koehler2012} demonstrated that aggregates made of amorphous silicates and carbons can hardly reach an opacity enhancement of 2.7 when 16 grains are stuck together, which already leads to very long coagulation timescales. To better understand the physical properties of L1506C, molecular gas emission lines ($^{13}$CO, C$^{18}$O, N$_2$H$^+$) were later observed and modelled by \citet{Pagani2010}. Explaining the molecular line profiles requires a central density at least seven times larger than the density inferred by \citet{Stepnik2003}. This discrepancy comes mostly from the low angular resolution of the PRONAOS data ($\sim 4$'). Furthermore, \citet{Pagani2010} observed a very strong CO depletion inside L1506C, up to a factor of 30 (lower limit) in the densest part of the filament, and an extremely low turbulence with an upper limit of $v_{FWHM} \sim 68$~m/s. They also showed the evidence of a collapsing core detached from its envelope, which would be the signature of a prestellar core in the making. Thus, L1506C appears to be very interesting because of both variations in dust properties and its key evolutionary stage. Finally, because L1506C is one of the most cited examples to illustrate grain growth in dense clouds, we revisit the dust emission analysis. We take advantage of the angular resolution of the Herschel data, which is ten times higher than the resolution of PRONAOS data, and of the constraints given by the molecular gas study \citep{Pagani2010}. We also expand our analysis to the full length of L1506, instead of considering just L1506C. Our aim is to check whether the dust optical properties have actually evolved in L1506 and if so, at which density and by how much.

The paper is organised as follows. In Sect. \ref{section_observations}, we describe the Herschel observations and data reduction. In Sect. \ref{section_extinction}, we explain how we produced visual extinction maps from the 2MASS data. In Sect. \ref{section_models}, we present the models we use to calculate the dust emission, extinction, and the radiative transfer through the filament. In Sect. \ref{section_modelling_fitting_methods}, we detail the methods that are used to simultaneously fit the dust emission in the five Herschel spectral bands and how the best-fitting models are compared to extinction. In Sect. \ref{section_modelling}, we give the results from modelling L1506 and then discuss them in Sect. \ref{section_discussion}. We give our conclusions in Sect. \ref{section_conclusions}.

\section{Herschel observations and data reduction}
\label{section_observations}

Within the framework of the two Herschel key programmes, ”Evolution of interstellar dust” \citep{Abergel2010} and the ”Herschel Gould Belt survey” \citep{Andre2010}, the Taurus S3 filament (L1506) has been mapped by the ESA Herschel Space Observatory with the SPIRE (Spectral and Photometric Imaging Receiver) and PACS (Photodetector Array Camera and Spectrometer) instruments \citep{Griffin2010, Poglitsch2010}. The observations were conducted on September 19, 2010 in the parallel-mode with fast, $60\arcsec$/sec, scanning speed. Two perpendicular maps were observed both with PACS (70 and 160~$\mu$m) and SPIRE (250, 350, and 500~$\mu$m) of a commonly covered $100\arcmin \times 60 \arcmin$ field for a total observing time of 2.3 hours.

The SPIRE maps reported in this paper are the Level 2 naive maps delivered by the Herschel Science Centre (HIPE version 7.0.1991) with standard corrections for instrumental effects and glitches. Striping induced by offsets in the flux calibration from one detector to another was removed using the Scan Map Destriper module included in the HIPE environment. The overall absolute flux accuracy is dominated by the calibration uncertainty and is conservatively estimated to be $\pm 7\%$.

The PACS data were processed with the map-maker SANEPIC (Signal and Noise Estimation Procedure Including Correlations), as described in \citet{Patanchon2008}. The overall absolute flux accuracy is dominated by the calibration uncertainty and is conservatively estimated to be $\pm 20\%$.

\begin{figure}[!ht]
\centerline{
\begin{tabular}{c}
\includegraphics[width=0.35\textwidth]{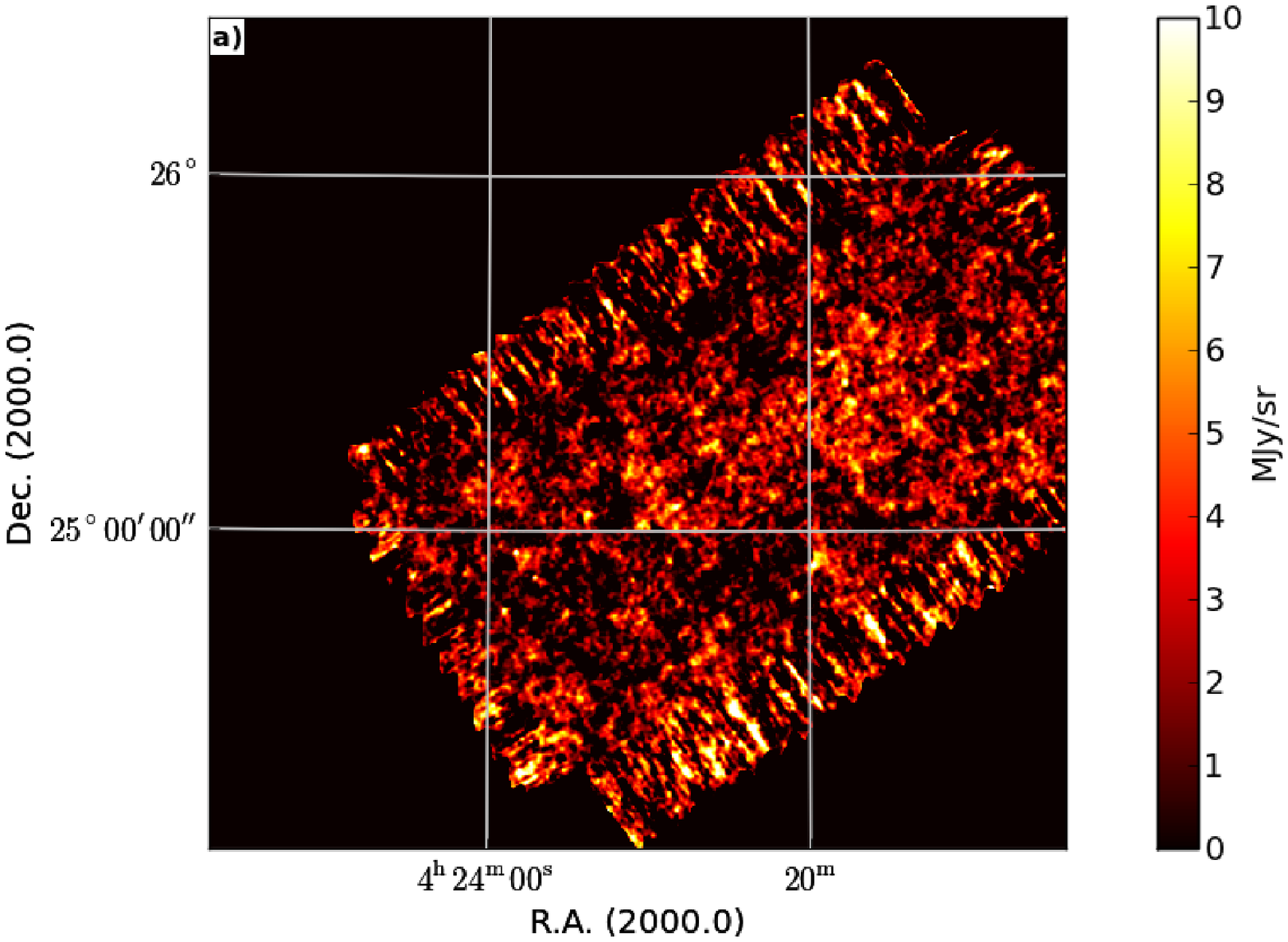} \\
\includegraphics[width=0.35\textwidth]{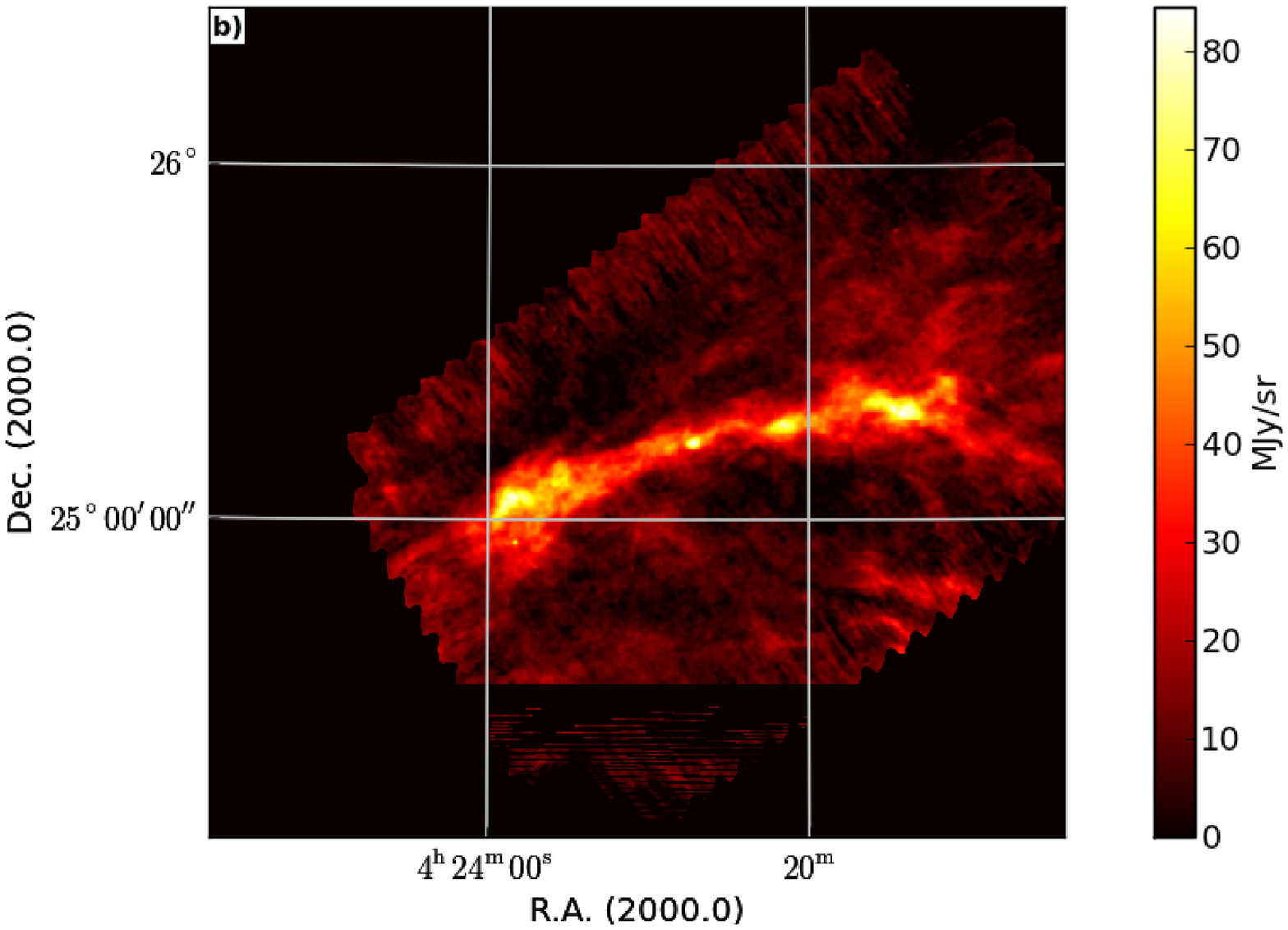} \\
\includegraphics[width=0.35\textwidth]{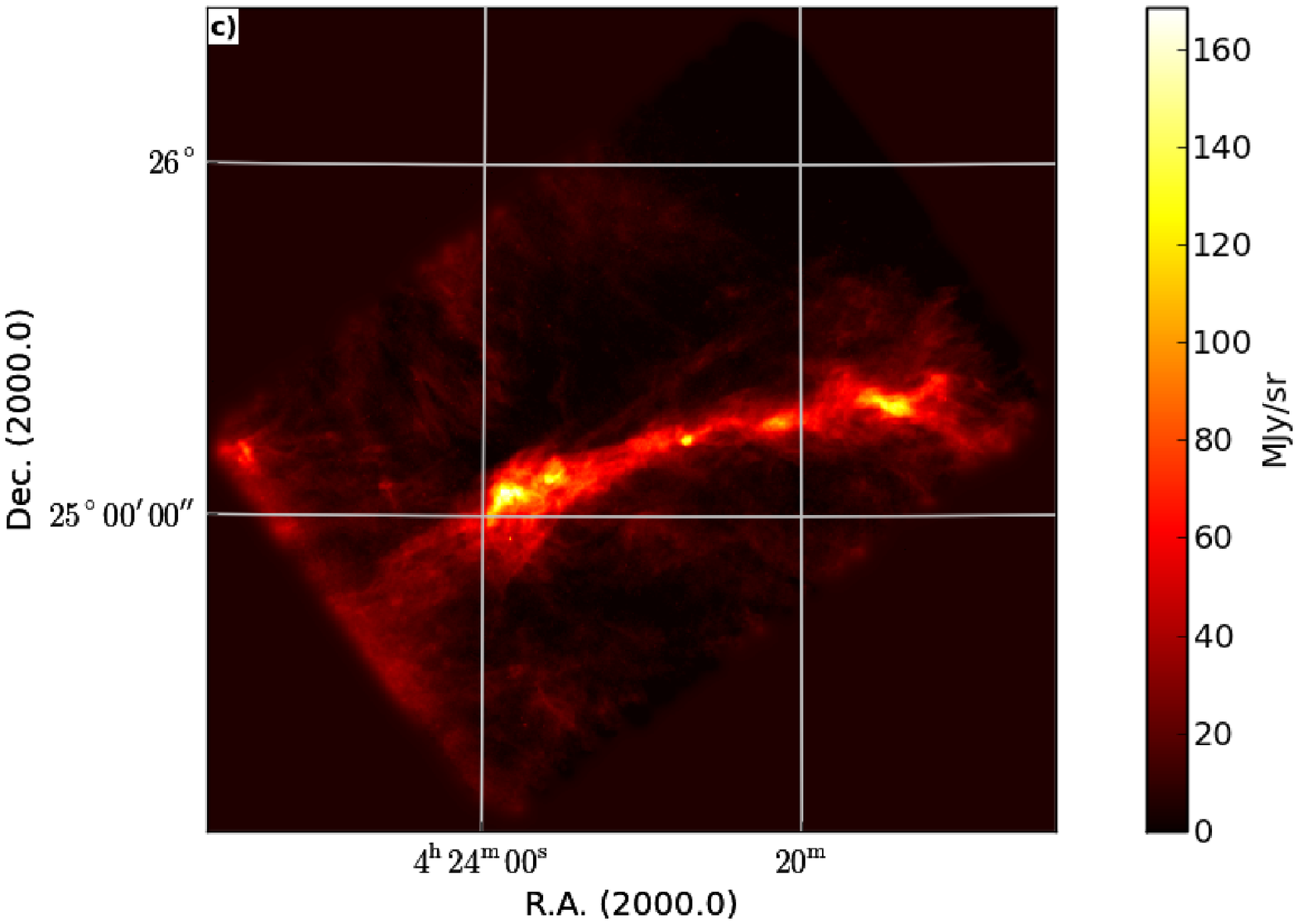} \\
\includegraphics[width=0.35\textwidth]{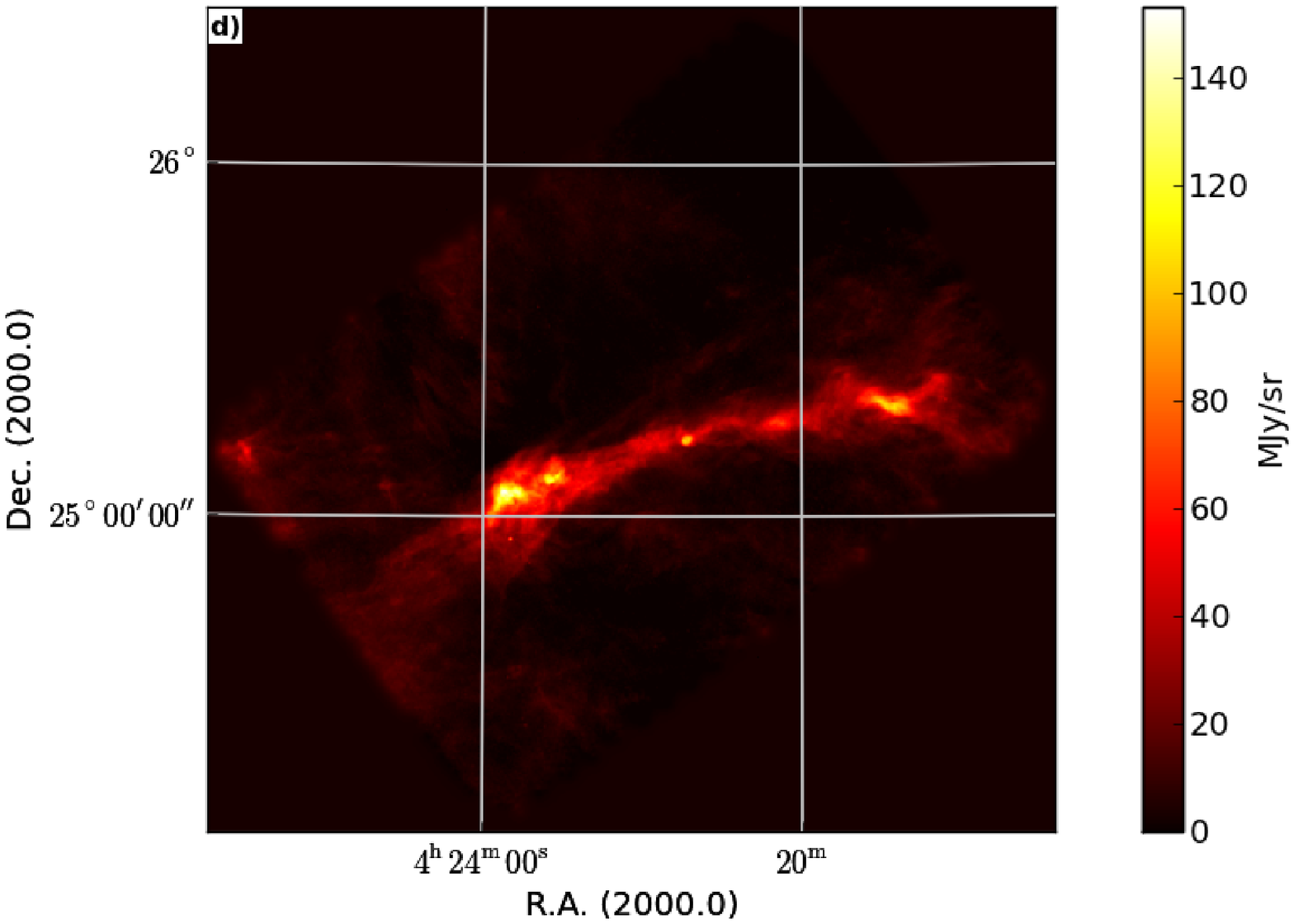} \\
\includegraphics[width=0.35\textwidth]{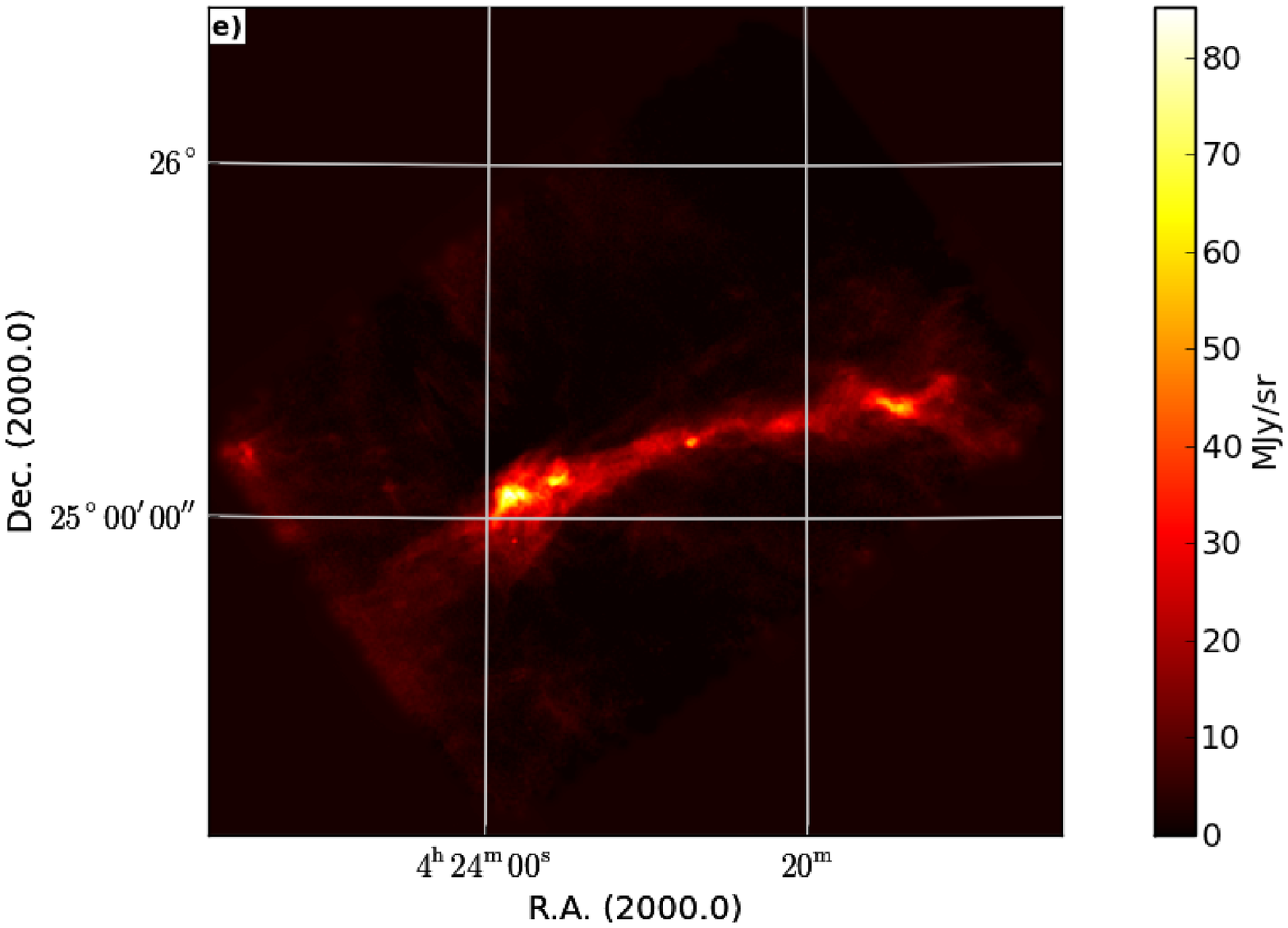}
\end{tabular}}
\caption{Brightness maps of the Taurus filament at 70 $\mu$m (a), 160 $\mu$m (b), 250 $\mu$m (c), 350 $\mu$m (d), and 500 $\mu$m (e).}
\label{fig_Herschel_maps} 
\end{figure}

The maps have an angular resolution of 8.4, 13.5, 18.2, 24.9, and 36.3\arcsec at 70, 160, 250, 350, and
500\,$\mu$m, respectively. The maps were all convolved to the angular resolution of the 500\,$\mu$m SPIRE band of 36.3\arcsec (0.02~pc at 140~pc), assuming Gaussian beams. The maps used in this paper are presented in Fig.\ref{fig_Herschel_maps}.

\section{Extinction from 2MASS data}
\label{section_extinction}

The extinction map of the L1506 filament is computed with the NICER method \citep{Lombardi2001}. We use the Two Micron All Sky Survey (2MASS) data \citep{Skrutskie2006} from which measurements of low photometric reliability were first removed ($S/N < 5$ or $\sigma > 0.217$~mag). The sigma-clipping algorithm described in \citet{Lombardi2001} is used with the default threshold of 3$\sigma$. Because the maximum $A_{{\rm V}}$ is only of about 10 to 20 magnitudes in L1506, we omit extinction-dependent colour corrections of the 2MASS data, since their effect would be only at the level of one percent \citep{Roy2013}. The calculations also require an assumption for the shape of the extinction curve of the dust grains. This curve is usually described with the parameter $R_{{\rm V}} = A_{\rm V} / E({\rm B}-{\rm V})$, which is expected to vary from $R_{{\rm V}} \sim 3.1$ in the diffuse medium to $R_{{\rm V}} \sim 5.5$ or even higher in dense molecular clouds \citep{Cardelli1989, Fitzpatrick1999}. The changes are however small in the near-IR domain \citep{Cardelli1989} and should not have a significant effect on results expressed as near-IR extinction\footnote{Recent work started to question this issue but without giving physical explanation for a E(J-K) variation \citep{vanBreemen2011}.}. However, a change from $R_{{\rm V}} = 3.1$ to $R_{{\rm V}} = 5.5$ corresponds to a $\sim 14$\% drop in the values expressed in $A_{{\rm V}}$. For this reason, we computed two different extinction maps: one with $R_{\rm V} = 3.1$ and the other with $R_{\rm V} = 5.5$, assuming that the extinction curve follows the prescription of \citet{Cardelli1989}. These two maps are calculated at a spatial resolution of FWHM=200\arcsec (i.e., 3.3\arcmin). In the case of a varying column density within individual cells, the NICER estimates can be biased towards low values \citep{Juvela2008}. As shown in Appendix A, this bias is of the order of 25\% for the densest parts of the filament and reaches $\sim 15$\% for the less dense ones.

\section{Models: Shape of the cloud and dust properties}
\label{section_models}

As the dust emission profiles are affected simultaneously by the radiative transfer and the dust properties, we use the coupling of the 3D Monte-Carlo radiative transfer code called CRT ({\it Continuum Radiative Transfer}, Juvela 2005) and the dust emission and extinction code DustEM \citep{Compiegne2011, Ysard2012}. The DustEM code is used to determine the dust temperature distribution and emission. It is a versatile code that works with any kind of dust properties and size distributions.

\subsection{Shape of the cloud}
\label{section_CRT}

As described in \citet{Juvela2005} and \citet{Ysard2012}, CRT is a 3D Monte-Carlo radiative transfer code that allows for three types of cloud geometry: spheres, cylinders, and three-dimension clouds defined on a Cartesian grid.

As can be seen in Fig. \ref{fig_Herschel_maps}, L1506 seems to have a filamentary structure. However, this may be the result of its projection onto the plane of the sky; alternatively, it could be a sheet, projected edge-on. To model the emission lines of $^{13}$CO, C$^{18}$O, and N$_2$H$^+$, \citet{Pagani2010} showed that L1506C must have a central H$_2$ density of $20\,000 \lesssim n(H_2) \lesssim 50\,000$~cm$^{-3}$. This leads to a peak column density of about $8 \times 10^{21} \lesssim N(H_2) \lesssim 2 \times 10^{22}$~H$_2$/cm$^2$ for a spherical model, which is similar to the results of \citet{Goldsmith2008} for the entire length of L1506. The ratio of the column density to the local density gives a rough estimate of the depth of the cloud in the direction perpendicular to the plane of the sky, which is about $0.05 \lesssim N(H_2) / n(H_2) \lesssim 0.32$~pc. This depth is comparable to the width of the cloud projected onto the plane of the sky, which is about 3\arcmin~to 5\arcmin~along the filament, equivalent to 0.12 to 0.3~pc at a distance of 140~pc. This width is also much smaller than the total length of L1506, which is about 2.5~pc. We conclude it is a good approximation to model L1506 as a cylinder-like filament.

Many observational studies have shown that the density distribution of molecular clouds appears to be flat at their centres and then decreases following a power-law at larger radii \citep{Ward-Thompson1994, Bergin2007}. Recently, Herschel data were used to characterise this power-law in the case of filamentary clouds. For instance, \citet{Arzoumanian2011} found that the 27 filaments observed in the IC5146 molecular cloud have density profiles $\sim r^{-2 \pm 0.5}$. A similar result was found by \citet{Palmeirim2013} in the Taurus molecular complex for the dense filaments B211 and L1495 and by \citet{Juvela2012} for 24 filaments identified as Galactic cold clumps with Planck HFI data \citep{PlanckMontier2011}. As a result, we represent the density distribution of L1506 with the following Plummer-like function:
\begin{equation}
n_H(r) = \frac{n_C}{1 + (r/R_{flat})^2},
\end{equation}
where $n_C$ is the central density and $R_{flat}$ is the central ``flat'' radius under which the density varies little. Except where otherwise stated, all the modelled filaments are assumed to lay in the plane of the sky with their axis of symmetry along L1506.

\subsection{Dust optical properties}
\label{section_dust}

\begin{figure}[!ht]
\centerline{
\includegraphics[width=0.35\textwidth]{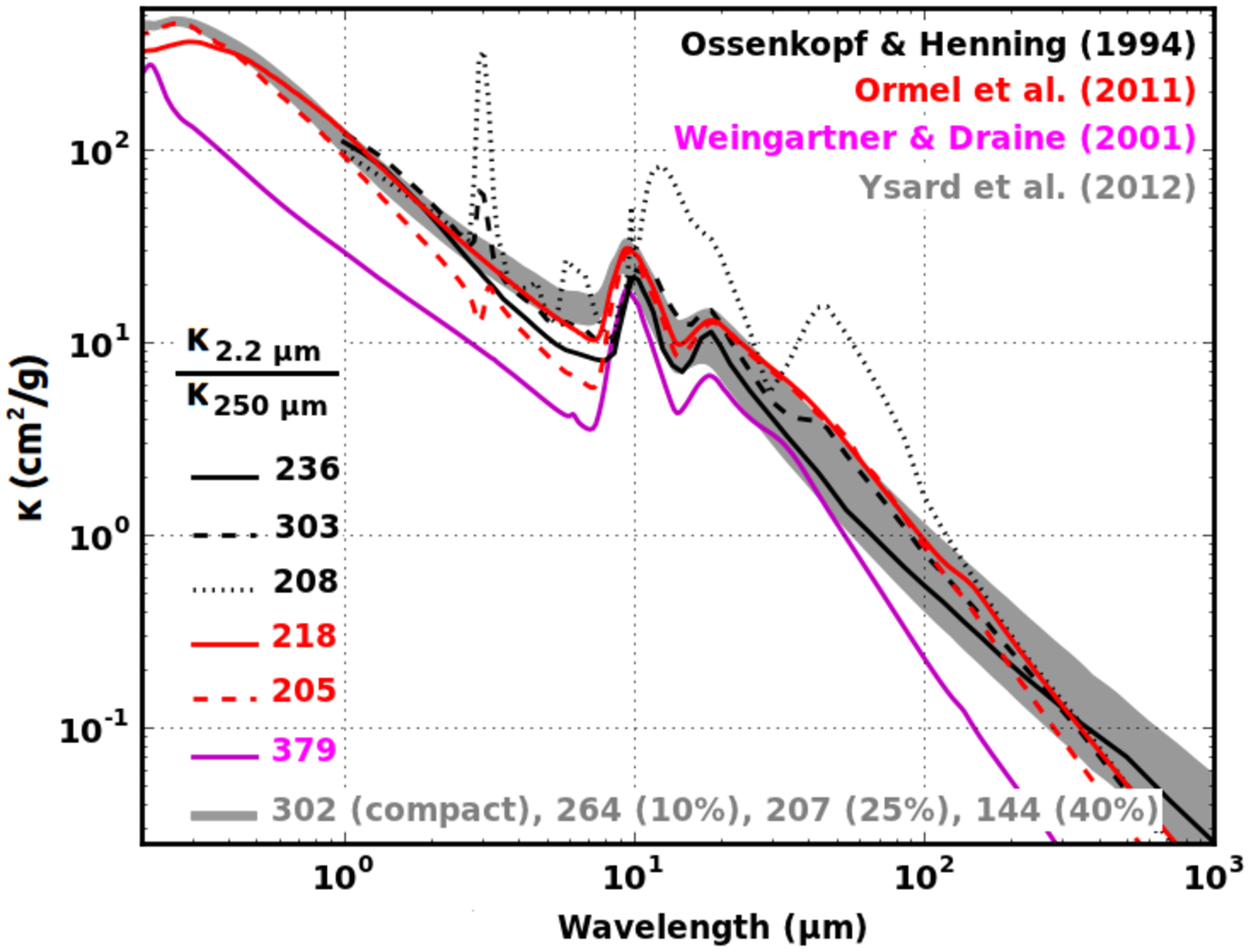}}
\caption{Opacity for the aggregate populations used in this paper (grey area). The magenta line shows the opacity of the dust model ``A'' of \citet{Weingartner2001}. We also display the opacity of the bare aggregates (black line) and of the aggregates covered by a thin (black dashed line) or thick (black dotted line) ice mantle described in \citet{Ossenkopf1994}. The red lines show the aggregate opacities computed by \citet{Ormel2011} without (solid line, their model {\it sil, gra}) and with an ice mantle (dashed line, their model {\it ic-sil, ic-gra}). The numbers on the left side of the figure give the NIR to far-IR opacity ratio for these dust models.}
\label{compare_kappa} 
\end{figure}

For the dust optical properties, we consider two of the cases studied by \citet{Ysard2012}. First, we consider a population of grains typical of the diffuse ISM as defined by \citet{Compiegne2011}: the DHGL populations (Dust at High Galactic Latitude). The DHGL populations consist of interstellar PAHs, small and large amorphous carbons (SamC and LamC, respectively), and astronomical silicates (aSil). For the small grains (PAHs \& SamC), we use log-normal size distributions with central radii $a_0$ and widths $\sigma$. For the larger grains (LamC \& aSil), we use a power-law distribution $a^{\alpha}$, starting at $a = 4$ nm, with an exponential cut-off $e^{-[(a-a_t)/a_c]^2}$ for $a \geq a_t$ \citep{Weingartner2001, Compiegne2011}. The dust model abundances and the parameters of the size distributions are given in Table 1. 

Second, we consider a population of ``evolved'' grains\footnote{The motivation behind the dust models described in this section is discussed further in \citet{Ysard2012}.}. These grains consist of aggregates, which are a mixture of small and large grains, with both carbons and silicates. Aggregates are expected to form in dense molecular clouds, possibly with large fractions of voids or porosity \citep{Dorschner1995}. Their optical properties are calculated using Mie theory combined with the effective medium theory (EMT) using the Bruggeman rule \citep{Bohren1983}. As in \citet{Ysard2012}, we consider compact aggregates and porous aggregates with 10\%, 25\%, and 40\% of voids. The dust model abundances and the parameters of the size distributions are presented in Table 1. Through aggregation, the dust opacity is increased by a factor equal to the ratio of the absorption efficiency of the aggregates to the mean absorption efficiency of isolated DHGL carbon and silicate grains of the same volume. This corresponds to an increase of 1.84 for compact aggregates, 2.16 for porous aggregates with 10\% of voids, 2.84 for 25\%, and 3.89 for 40\%. These aggregates have a low opacity spectral index in the far-IR and the submm of $\beta \sim 1.3$, resulting from the shape of the absorption efficiency of the large grains of the DHGL populations (see also Sect. \ref{section_caveats}). The aggregate opacities are displayed in Fig. \ref{compare_kappa}. For comparison, the aggregate dust models, as obtained after $10^5$~yrs of coagulation, of \citet{Ormel2011} and \citet{Ossenkopf1994} for $n_H = 10^6$~H/cm$^3$ are shown in this figure. We also show the Milky Way dust model ``A'' of \citet{Weingartner2001}, which is not composed of aggregates but of diffuse-ISM type grains \citep{Li2001} with a modified size distribution to allow for bigger grains. This model was normalized according to \citet{Draine2003} and corresponds to $R_V = 5.5$. We note that the opacity of our model of aggregates is close to those developed by \citet{Ormel2011} and \citet{Ossenkopf1994}. The main difference between these models is their spectral index in the far-IR and submm.

\begin{table}
\label{tableau_dustem}
\centering
\caption{Dust model abundances and size distribution parameters (see Sect. \ref{section_dust} for details). The parameter $\rho$ is the grain mass density, $Y$ is the mass abundance per H, $\kappa_{250 \; \mu{\rm m}}$ is the opacity at 250 $\mu$m, and $\beta$ is the intrinsic opacity spectral index for $\lambda > 100 \; \mu$m for each dust population. For aggregates, the percentages correspond to the volume of voids they contain, which indicates how porous they are.}
\begin{tabular}{ccccccc}
\hline
\hline
\multicolumn{7}{c}{Small grains (DHGL)} \\
\hline
     & $\rho$     & $\sigma$ & $a_0$ & $Y$                   & $\kappa_{250 \; \mu{\rm m}}$ & $\beta$\\
     & (g/cm$^3$) &          & (nm)  & ($M_{dust}/M_H$)      & (cm$^2$/g)                   &        \\
PAH  & 2.24       & 0.40     & 0.64  & 7.80$\times 10^{-4}$  & 0.001                        &        \\
SamC & 1.81       & 0.35     & 2.00  & 1.65$\times 10^{-4}$  & 0.002                        & 1.55   \\
\hline
\hline
\multicolumn{7}{c}{Big grains (DHGL)} \\
\hline
     & $\rho$ & $\alpha$ & $a_c, a_t$ & $Y$                  & $\kappa_{250 \; \mu{\rm m}}$ & $\beta$ \\
     &        &          & (nm)       &                      &                              &         \\
LamC & 1.81   & -2.8     & 150.0      & 1.45$\times 10^{-3}$ & 0.014                        & 1.55    \\
aSil & 3.5    & -3.4     & 200.0      & 7.8$\times 10^{-3}$  & 0.034                        & 2.11    \\
\hline
\hline
\multicolumn{7}{c}{Aggregates} \\
\hline
     & $\rho$ & $\alpha$ & $a_c, a_t$ & $Y$                  & $\kappa_{250 \; \mu{\rm m}}$ & $\beta$  \\
0\%  & 2.87   & -2.4     & 234.0      & 1.02$\times 10^{-2}$ & 0.111                        & 1.33     \\
10\% & 2.59   & -2.4     & 242.0      & 1.02$\times 10^{-2}$ & 0.140                        & 1.32     \\
25\% & 2.16   & -2.4     & 256.0      & 1.02$\times 10^{-2}$ & 0.208                        & 1.30     \\
40\% & 1.72   & -2.4     & 276.0      & 1.02$\times 10^{-2}$ & 0.331                        & 1.27     \\
\hline
\end{tabular}
\end{table}

\section{Modelling and fitting methods}
\label{section_modelling_fitting_methods}

In this section, we describe how we subdivide the observed irregular filament into four separate and homogeneous filamentary pieces to probe the dust properties at various locations along L1506. We also detail our fitting method and the free parameters considered.

\subsection{Filament description}
\label{section_preparation}

\begin{figure*}[!ht]
\centerline{
\begin{tabular}{cc}
\includegraphics[width=0.45\textwidth]{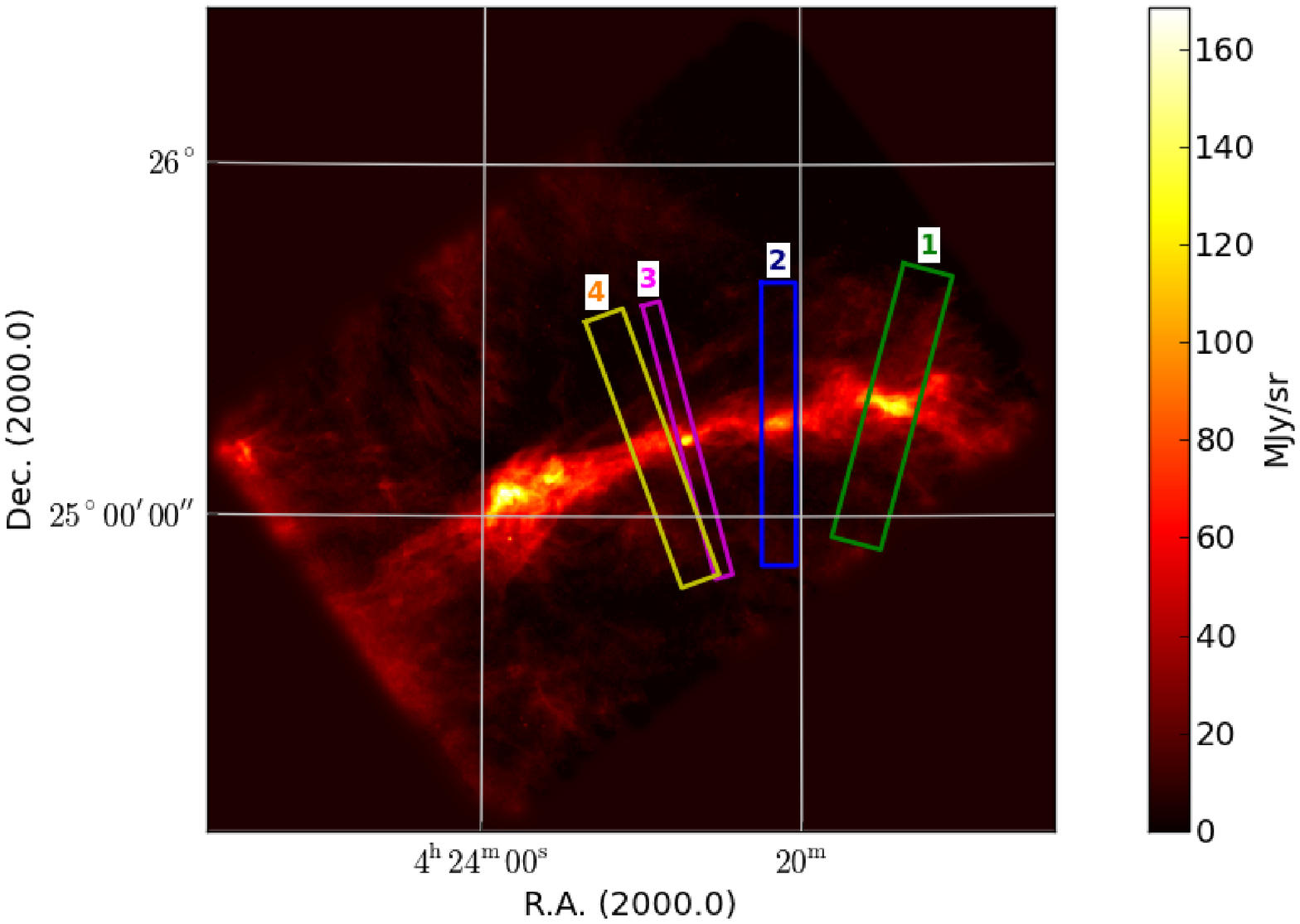} & \includegraphics[width=0.45\textwidth]{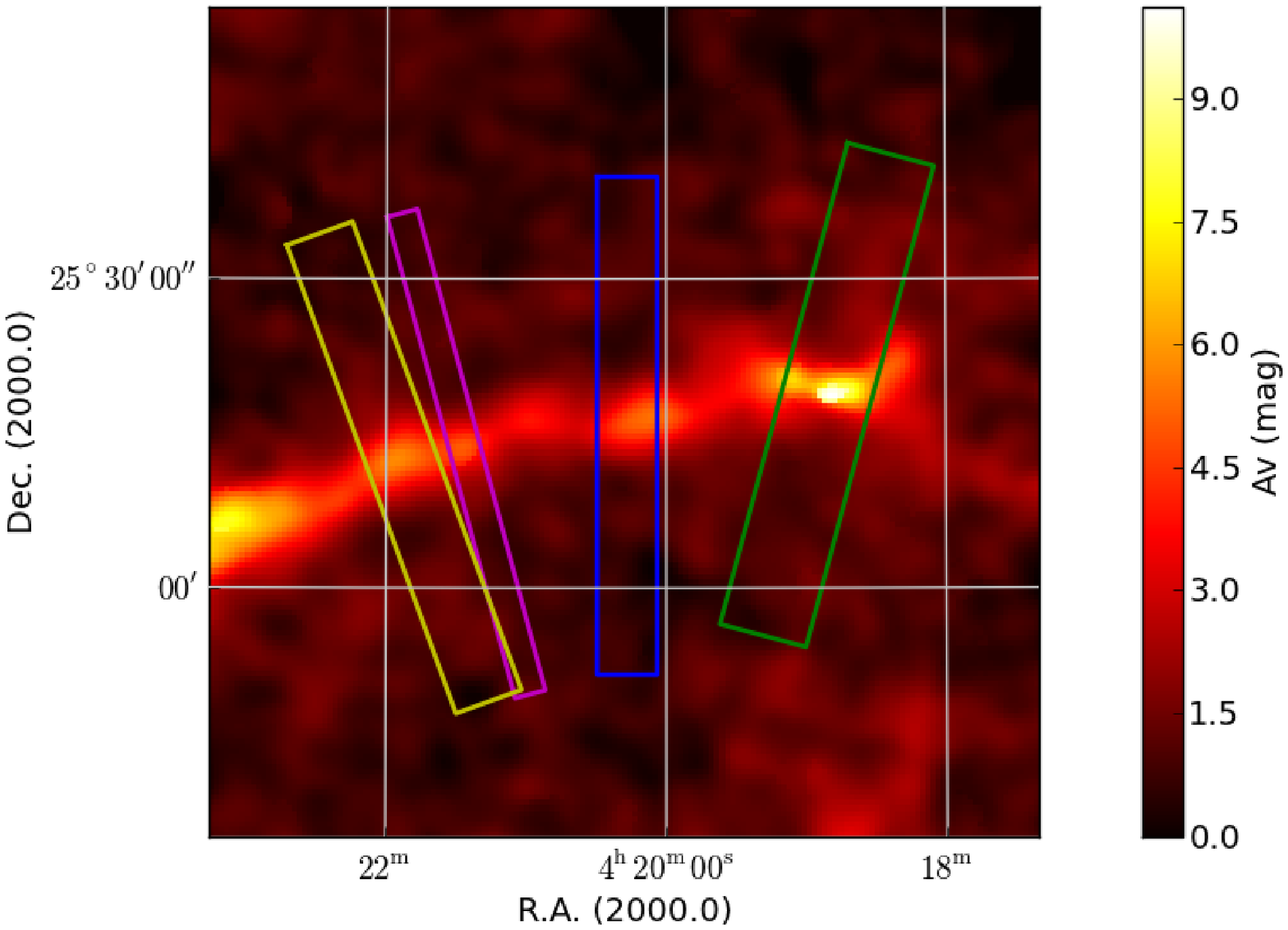} \\
\end{tabular}}
\caption{{\it Left}: Brightness map of the filament at 250~$\mu$m. The rectangles show the positions of the four cuts of Table 2 (green is the first one, blue the second, magenta the third, and yellow the fourth). {\it Right}: Extinction map $A_{{\rm V}}$ at 0.55~$\mu$m calculated from 2MASS data for $R_{{\rm V}} = 3.1$.}
\label{figures_profiles} 
\end{figure*}

\begin{figure*}[!ht]
\centerline{
\includegraphics[width=0.92\textwidth]{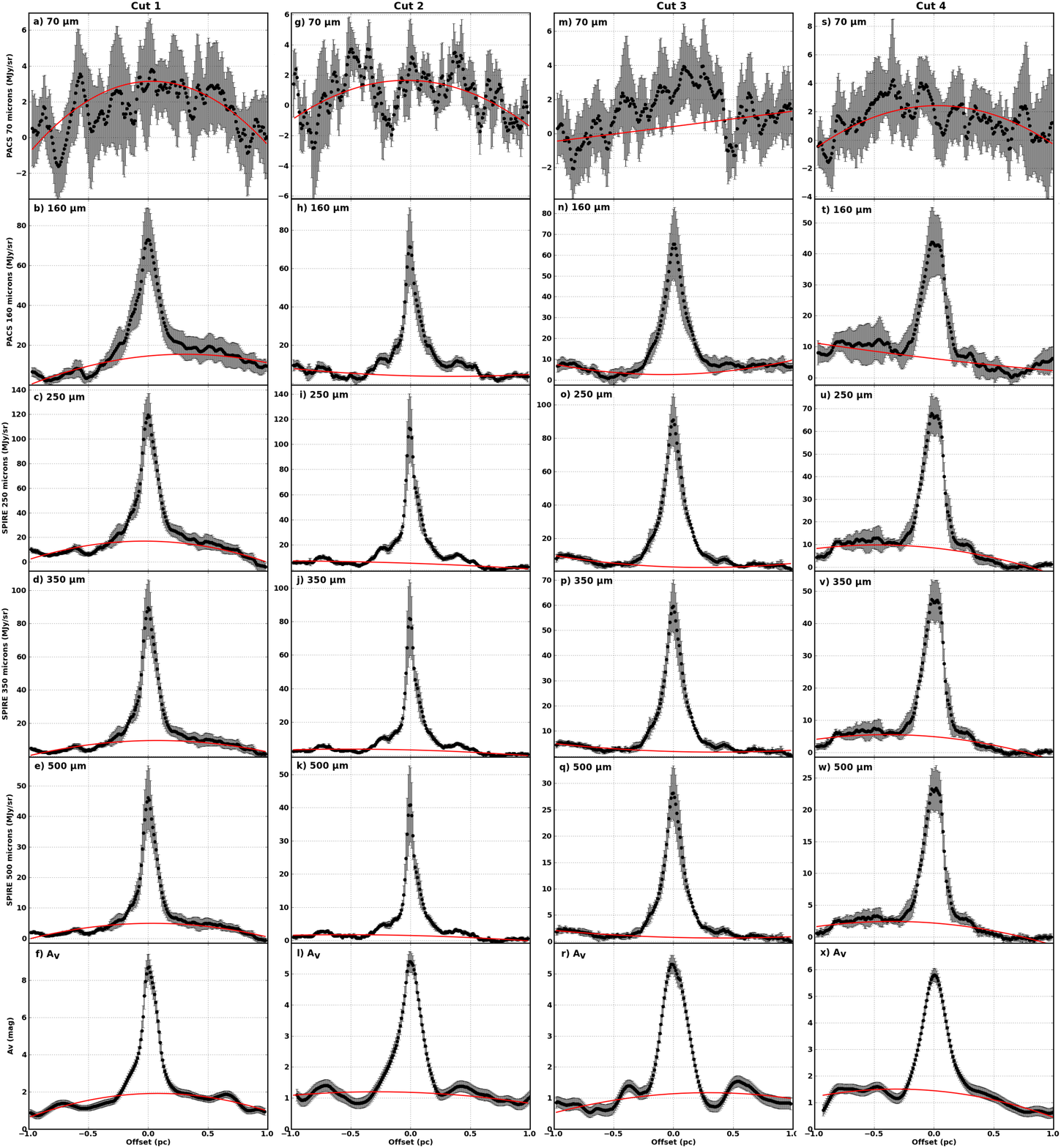}}
\caption{Emission brightness profiles of the first cut (see Table 2) at 70 $\mu$m (a), 160 $\mu$m (b), 250 $\mu$m (c), 350 $\mu$m (d), and 500 $\mu$m (e), and extinction profile calculated from the 2MASS data with $R_V = 3.1$ (f). The same for the second cut (g-l), the third cut (m-r), and the fourth cut (s-x). The red lines show the second order polynomials fitted to remove the relatively diffuse envelope from the densest filament.}
\label{fig_emission} 
\end{figure*}

We selected four locations along the filament that we want to model (Table 2 and Fig. \ref{figures_profiles}). The first location, or first cut across the filament, corresponds to the filament modelled by \citet{Stepnik2003} and \citet{Pagani2010} that they named L1506C (green box in Fig. \ref{figures_profiles}). The second and the fourth cuts cross less dense areas, while the third one is located on a brighter and more compact spot (blue, yellow, and magenta boxes in Fig. \ref{figures_profiles}, respectively). To define the brightness profiles corresponding to one location, we use a box perpendicular to the filament, the width of which is chosen so that the brightness along the filament remains approximately constant (Fig. \ref{figures_profiles}). Then, the brightness profile is defined as the average value of each line of pixels parallel to the filament axis. The error bars are defined as the quadratic addition of the standard deviation of each line of pixels plus 7\% of the brightness profile in the case of SPIRE bands and 20\% in the case of PACS bands to account for the calibration uncertainties (see Sect. \ref{section_observations}). The corresponding profiles are shown in Fig. \ref{fig_emission} for the emission and the visual extinction. We note that the resolution of 36.3\arcsec~allows us to resolve the filament.

As we are mostly interested in the densest component of the filament where the gas is predominantly molecular, we separate the filament into two parts: a diffuse envelope and the actual dense molecular filament. The diffuse envelope is fitted with a second order polynomial as in \citet{Stepnik2003} for offsets of 0.5 to 1~pc from the centre of the cut coordinates given in Table 2. The envelopes in emission and extinction are shown in Figs. \ref{fig_emission}. The actual brightness and extinction profiles that we fit in the following sections are what remains when the polynomials are removed from the data. Using this subtraction has also the advantage of defining a zero-level common to all the SPIRE and PACS channels and to the extinction map.

\begin{table}
\label{table_cuts}
\centering
\caption{Coordinates of the centre and widths of the cuts across the L1506 filament.}
\begin{tabular}{lccc}
\hline
\hline
           & Ra (J2000) & Dec (J2000) & Width \\
\hline
Cut 1 & 4h18m51s   & +25d18m43.6s     & 8.6\arcmin / 0.35 pc\\
Cut 2 & 4h20m16.7s & +25d15m46.3s     & 3.0\arcmin / 0.12 pc \\
Cut 3 & 4h21m25.7s & +25d13m2.5s      & 5.8\arcmin / 0.24 pc\\
Cut 4 & 4h21m53.2s & +25d11m41s       & 8.2\arcmin / 0.33 pc \\
\hline
\end{tabular}
\end{table}

\subsection{Fitting method}
\label{section_method}

For each part of the filament, our aim is to simultaneously fit the five Herschel brightness profiles and the 2MASS extinction profile. For a given grain type (size distribution and optical properties), the parameters to be fitted are the incident radiation field, the outer radius of the filament $R_{out}$, its flat central radius $R_{flat}$, and its central density $n_C$. For the grains, we use the properties described in Sect. \ref{section_dust}, either mixed or not. \citet{PlanckAbergel2011} computed the dust grain temperature in the Taurus molecular complex, performing modified blackbody fits of the Planck-HFI and IRAS data (100~$\mu$m to 2~mm). In the diffuse medium surrounding the dense filaments, the thermal dust temperature ranges between $\sim$16 to 20~K. These values are close to the value measured in the standard diffuse ISM \citep{PlanckMAMD2011, PlanckBernard2011}. The radiation field illuminating the L1506 filament is thus assumed to be close to the standard radiation field, hereafter addressed as the ISRF \citep{Draine1985}. As a result, we test several radiation fields: the ISRF and this same ISRF extinguished and reddened by a layer of DHGL grains with $A_{{\rm V}}^{ext} = 0.05, 0.1, 0.2, 0.3, 0.4, 0.5, 0.6, 0.7, 0.8, 0.9$, and 1$^{\rm m}$ (corresponding roughly to $0.9 \gtrsim G_0 \gtrsim 0.3$). For $R_{out}$, we test values between 0.15 and 0.5~pc, and we try 30 different values of $R_{flat}$ between $R_{out}/200$ and $R_{out}/1.2$, which correspond to steps of 0.005~pc.

For each run, we take one set of those values (grain type, $A_{{\rm V}}^{ext}$, $R_{out}$, and $R_{flat}$), and the free parameter is the central density $n_C$ that we vary to minimize the $\chi^{2}$ for the fit between the Herschel and the modelled brightness profiles. Starting from the smallest $\chi^2$, we consider all the models for which the modelled profiles are within the error bars for the five Herschel bands. These models are then compared to the extinction profile calculated with the 2MASS data either for $R_{{\rm V}} = 3.1$ or 5.5 (see Sect. \ref{section_extinction}). For each model, if the modelled visual extinction at the centre of the modelled filament does not match the 2MASS $A_{{\rm V}}$ value, the model is discarded. The uncertainties on the derived physical parameters that are quoted in the following section reflect the size of the family of models, which provide a good fit to both the emission and the extinction profiles.

\section{Dust emission and extinction modelling}
\label{section_modelling}

We now apply the methods described in the previous section to the four cuts given in Table 2.

In all of the cases presented in this section, the best fits were obtained for the standard ISRF, so we do not present results obtained for attenuated radiation fields. Furthermore, the best fits were obtained for the same outer radius, irrespective of which kind of grains were considered. Consequently, we only show the results for those $R_{out}$ values, which are 0.3~pc for the first cut, 0.4~pc for the second cut, 0.32~pc for the third cut, and 0.2~pc for the fourth cut.

\subsection{DHGL}
\label{section_DHGL}

\begin{figure*}[!ht]
\centerline{
\includegraphics[width=0.92\textwidth]{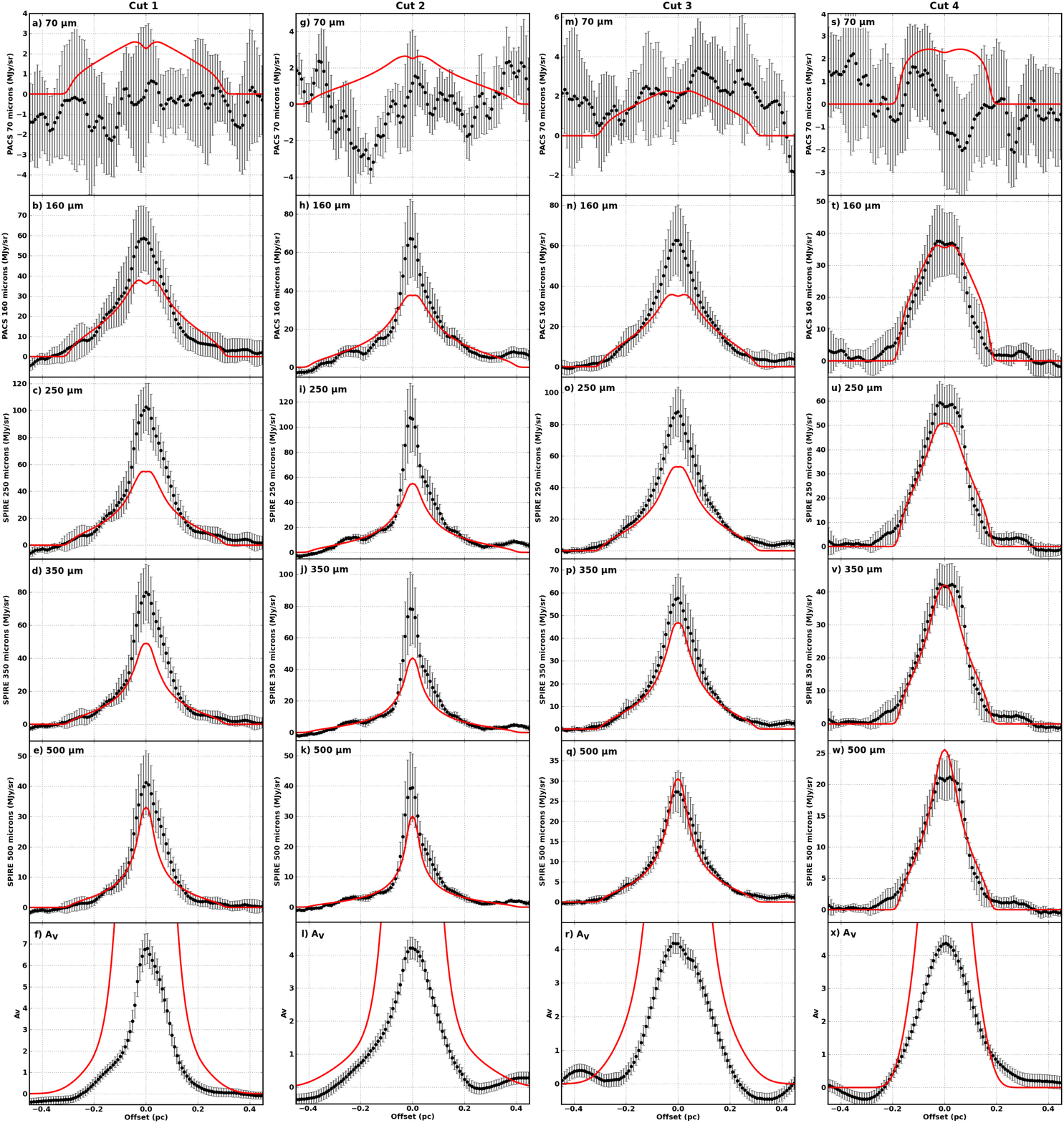}}
\caption{The red lines show the best fits with DHGL grains of the emission brightness profiles (background removed) for the first cut at 70 $\mu$m (a), 160 $\mu$m (b), 250 $\mu$m (c), 350 $\mu$m (d), and 500 $\mu$m (e), and the corresponding visual extinction profile (f). In the last row, the black dots show the extinction profile calculated from 2MASS data with $R_V = 3.1$. The same are shown for the second cut (g-l), the third cut (m-r), and the fourth cut (s-x).}
\label{fig_DHGL} 
\end{figure*}

First, we assume that the grain size distribution and optical properties in the dense filament are the same as the diffuse ISM. To mimic that, we attempt to fit the brightness profiles with cylinders filled with DHGL populations (see Sect. \ref{section_dust}). The best-fitting models are shown in Fig. \ref{fig_DHGL}, where it can be seen that the brightness profiles cannot be reproduced with DHGL populations for cut 1, 2, and 3. The emission profiles corresponding to the best-fitting models are within the error bars of the five Herschel emission profiles for only cut 4 (Fig. \ref{fig_DHGL}s-w), which, as can be seen in Fig. \ref{figures_profiles}, has the lowest density of any of the cuts. However, the extinction profile corresponding to this best fit model is a factor of 6.1 higher than the 2MASS extinction profile calculated for $R_{{\rm V}} = 3.1$, the typical value for dust in the diffuse medium (Fig. \ref{fig_DHGL}f, l, r, x). Factors of 4.4, 5.5, and 3.8 are found for cuts 1, 2, and 3. Indeed, the fitting of the brightness profiles requires high gas densities for the grains to be cold enough not to emit too much in the PACS 70~$\mu$m spectral band, leading to extinction values always higher than the 2MASS extinction. That means that the grain optical properties are not the same as the diffuse ISM. The grains have to evolve from the diffuse to the dense medium; this also applies to the relative abundance of the small grains (both large (LamC, aSil) and small grains (SamC) contribute to the emission at 70~$\mu$m).

To ascertain that this result does not depend on the choice of dust model, we repeated the fitting using the dust properties defined by \citet{Draine2007} for the diffuse ISM. This model is made of a mixture of amorphous silicates, graphite, and interstellar PAHs. Using this second dust model, we find similar results; that is, it is not possible to fit all the emission and extinction profiles simultaneously. We also considered the case of modelled cylinders with inclination angles of 10, 20, 30, and 40$\degr$, which did not permit to fit the data either. Consequently, we definitely exclude the possibility that the dust in L1506 is similar to that found in the diffuse ISM.

\subsection{Aggregates}
\label{section_aggregates}

\begin{figure*}[!ht]
\centerline{
\includegraphics[width=0.92\textwidth]{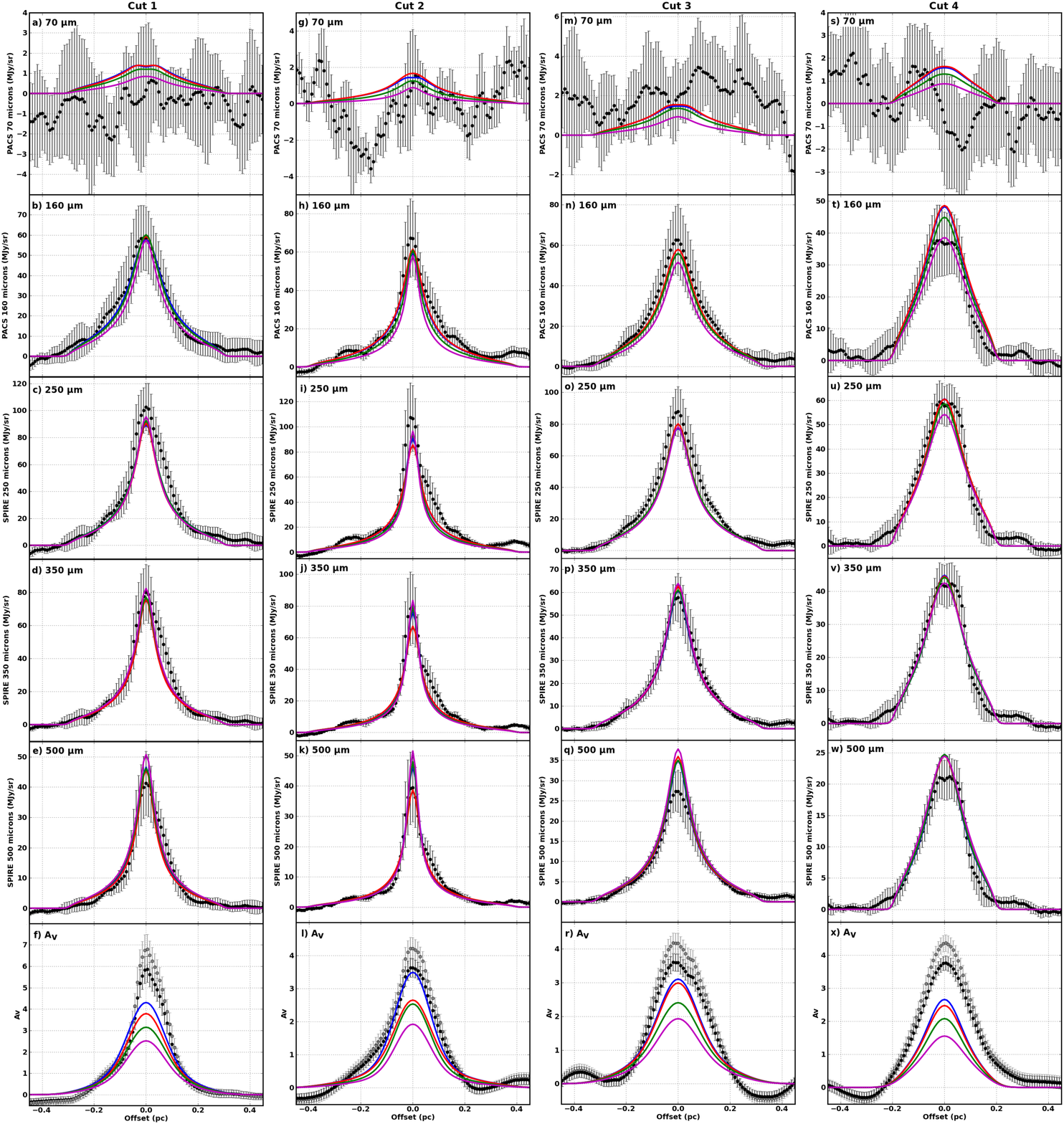}}
\caption{The lines show the best fits with aggregates of the emission brightness profiles for the first cut at 70 $\mu$m (a), 160 $\mu$m (b), 250 $\mu$m (c), 350 $\mu$m (d), 500 $\mu$m (e), and the corresponding extinction profile (f). The grey dots show the extinction profile calculated from 2MASS with $R_V = 3.1$ and the black ones with $R_V = 5.5$. The same are shown for the second cut (g-l), the third cut (m-r), and the fourth cut (s-x). The blue lines are for compact aggregates, the red lines for aggregates with 10\% of voids, the green lines with 25\% of voids, and the magenta lines with 40\% of voids.}
\label{fig_aggregates} 
\end{figure*}

As it appears that the grain properties have to vary from the diffuse surroundings of the filament to its centre, we model cylinders filled only with aggregates, which range from compact to porous grains containing 10, 25, or 40\% of voids. These aggregates are a mixture of all the dust populations of the DHGL model \citep{Compiegne2011}. In this case, we assume that all the grains are already evolved inside the filament, which no longer contains small grains, since they have already been incorporated into the aggregates. The best-fitting models are displayed in Fig. \ref{fig_aggregates}. The five brightness profiles of cut 1, 2, and 4 can be well reproduced with those aggregates. For cut 3, the SPIRE 500~$\mu$m brightness profile cannot be fitted by any of the aggregates tested here. Our best-fitting model for this cut leads to an overestimate of the emission at 500~$\mu$m by a factor of 1.3-1.4 depending on the aggregate type. In Fig. \ref{fig_aggregates}f, l, r, x, the extinction profiles corresponding to these best-fitting models are compared to the 2MASS extinction profiles calculated with $R_{{\rm V}} = 3.1$ and 5.5. Only the extinction profile of the second cut is well fitted by a cylinder filled with compact aggregates. For the other cuts, as the aggregates cool efficiently, the fits of the submillimetre emission require densities that are too low to explain the 2MASS visual extinction by factors of 1.5 to 2.5, 1.3 to 2.0, and 1.5 to 2.6 for cut 1, 3, and 4, respectively, and for compact to porous aggregates with 40\% of voids, respectively. Thus, considering a filament with an inclination angle of 10, 20, 30, or 40$\degr$ does not permit to fit the data either. This means that for at least three of the four cuts the grain properties have to vary {\it inside} the dense molecular part of the filament.

\subsection{Evolution across the filament}
\label{section_evolution}

The results of the previous sections indicate that the grain properties need to vary as a function of position within the filament, especially in its dense centre. In this case, we assume that DHGL populations are present in the outer layers and that the grain properties no longer correspond to DHGL but instead to aggregates above a given threshold density. We choose to represent this threshold by a step function to keep a reasonable number of free parameters. Thus, the value of the threshold density simply tells above which density the aggregates dominate the dust mass. This threshold density, $n_T$, is a new parameter and we test $n_T~=~125$, 250, 375, 500, 625, 750, 875, 1\,000, 1\,125, 1\,250, 1\,375, 1\,500, 1\,750, 2\,250, 2\,500, 3\,000, 3\,500, 4\,000, 4\,500, 5\,000, 5\,500, 6\,000, 6\,500, 7\,000, 7\,500, 8\,000, 8\,500, 9\,000, 9\,500, and 10\,000 H/cm$^3$. The fitting method remains the same: for a given parameter combination [grain type, $A_{{\rm V}}^{ext}$, $R_{out}$, $R_{flat}$, $n_T$], the free parameter to minimize the $\chi^{2}$ is still the central density $n_C$. The parameters of the best-fitting models are given in Table 3 and described in the following sections.

\subsubsection{First cut (L1506C)}
\label{section_first_cut}

\begin{figure}
\centerline{
\includegraphics[width=0.3\textwidth]{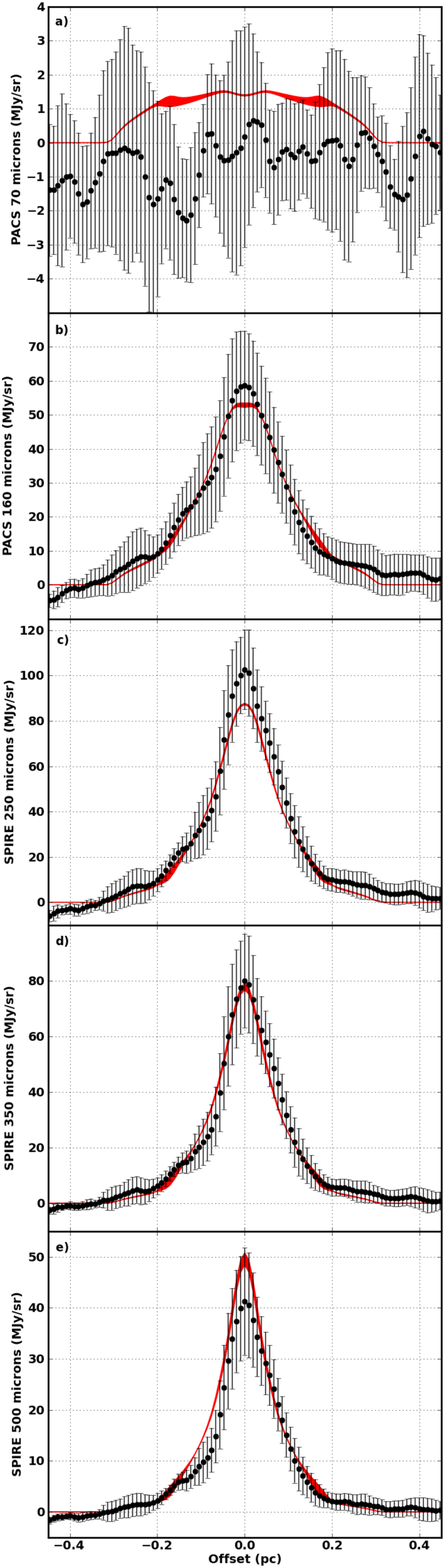}}
\caption{Brightness profiles of the first cut at 70, 160, 250, 350, and 500~$\mu$m (panel a, b, c, d, and e respectively). The red areas are the best-fitting models for porous aggregates with 10\% of voids and $n_T = 1\,500 \pm 250$~H/cm$^3$.}
\label{filament1_emission}
\end{figure}

\begin{figure}
\centerline{
\includegraphics[width=0.3\textwidth]{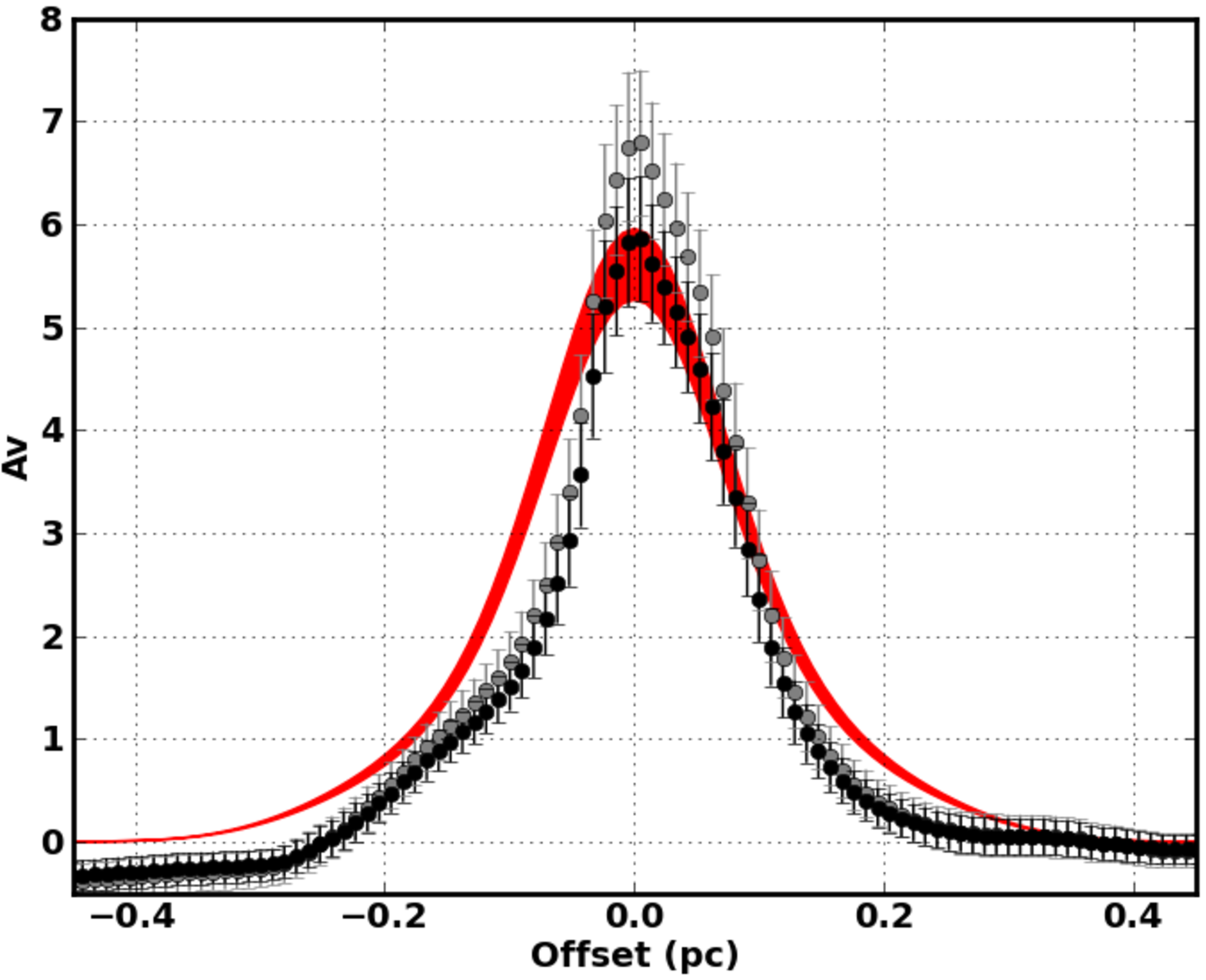}}
\caption{The grey dots trace the extinction profile of the first cut calculated from 2MASS data with $R_{{\rm V}}=3.1$ and the black ones with $R_{{\rm V}}=5.5$. The red area corresponds to the best-fitting models presented in Fig. \ref{filament1_emission}.}
\label{filament1_extinction} 
\end{figure}

The results for the first cut are shown in Figs. \ref{filament1_emission} and \ref{filament1_extinction}. The molecular gas lines in L1506C have been studied in detail by \citet{Pagani2010} and yield a strong constraint on the central density $n_C$. To explain the emission lines of N$_2$H$^+$, C$^{18}$O, $^{13}$CO, and C$^{17}$O, they found that $40\,000 \lesssim n_C \lesssim 100\,000$~H/cm$^3$. The brightness and extinction profiles can be fitted simultaneously with compact aggregates for $n_T = 875 \pm 125$~H/cm$^3$ or with porous aggregates with 10\% of voids for $n_T = 1\,500 \pm 250$~H/cm$^3$. However, in the case of compact aggregates the best-fitting models demand central densities between 150\,000 and 160\,000~H/cm$^3$. These values are too high when compared with the results of \citet{Pagani2010}, so we exclude compact aggregates. In the case of porous aggregates with 10\% of voids, the central densities are $90\,000 \lesssim n_C \lesssim 110\,000$~H/cm$^3$, which agrees with the densities required by the model explaining the gas emission lines. Therefore, as written previously, our best-fitting model requires the aggregates with 10\% of voids to prevail for $n_T = 1\,500 \pm 250$~H/cm$^3$, which corresponds to an offset of $0.19 \pm 0.03$~pc from the centre of the filament. It also corresponds to line-of-sights with a total $A_{{\rm V}} \sim 2.0-3.2$ (without envelope subtraction, Fig. \ref{fig_emission}f). This value roughly matches the threshold value of $A_{{\rm V}} = 3.3 \pm 0.1$ for the detection of ice in the Taurus dark clouds measured by \citet{Whittet1988}. Our threshold for the predominance of the aggregates coincides with the beginning of the decrease in the $^{13}$CO abundance observed by \citet{Pagani2010} between 0.15 and 0.23~pc. The desorption of volatile ice mantles can happen after cosmic-ray impacts \citep{Roberts2007}. However, this desorption is less efficient when the molecules have frozen-out onto the surface of large grains, the temperature of which is more stable while undergoing cosmic-ray impacts. Thus, the match between the start of the depletion of CO and the presence of large aggregates seems plausible. Furthermore, the presence of ice mantles on top of the grains is expected to increase the formation efficiency of the aggregates \citep{Ormel2009}.

\subsubsection{Second and fourth cuts}
\label{section_second_fourth_cuts}

\begin{figure}
\centerline{
\includegraphics[width=0.3\textwidth]{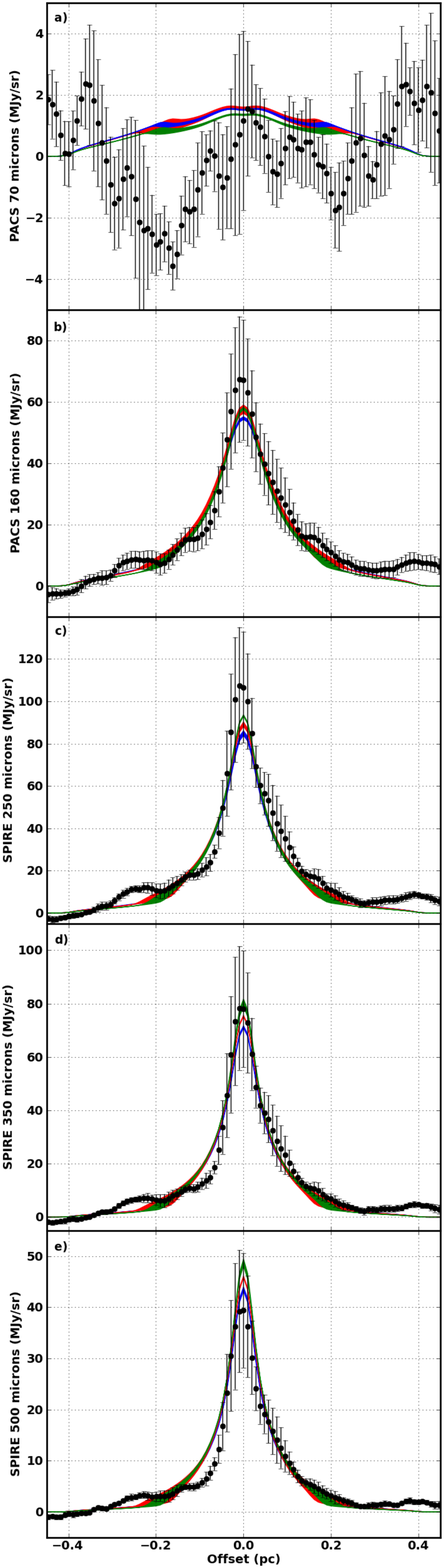}}
\caption{Brightness profiles of the second cut at 70, 160, 250, 350, and 500~$\mu$m (panel a, b, c, d, and e respectively). The blue areas are the best-fitting models for compact aggregates and $n_T=875 \pm 125$~H/cm$^3$, the red ones for porous aggregates with 10\% of voids and $n_T = 750 \pm 250$~H/cm$^3$, and the green ones for porous aggregates with 25\% of voids and $n_T = 625 \pm 125$~H/cm$^3$.}
\label{filament2_emission}
\end{figure}

\begin{figure}
\centerline{
\includegraphics[width=0.3\textwidth]{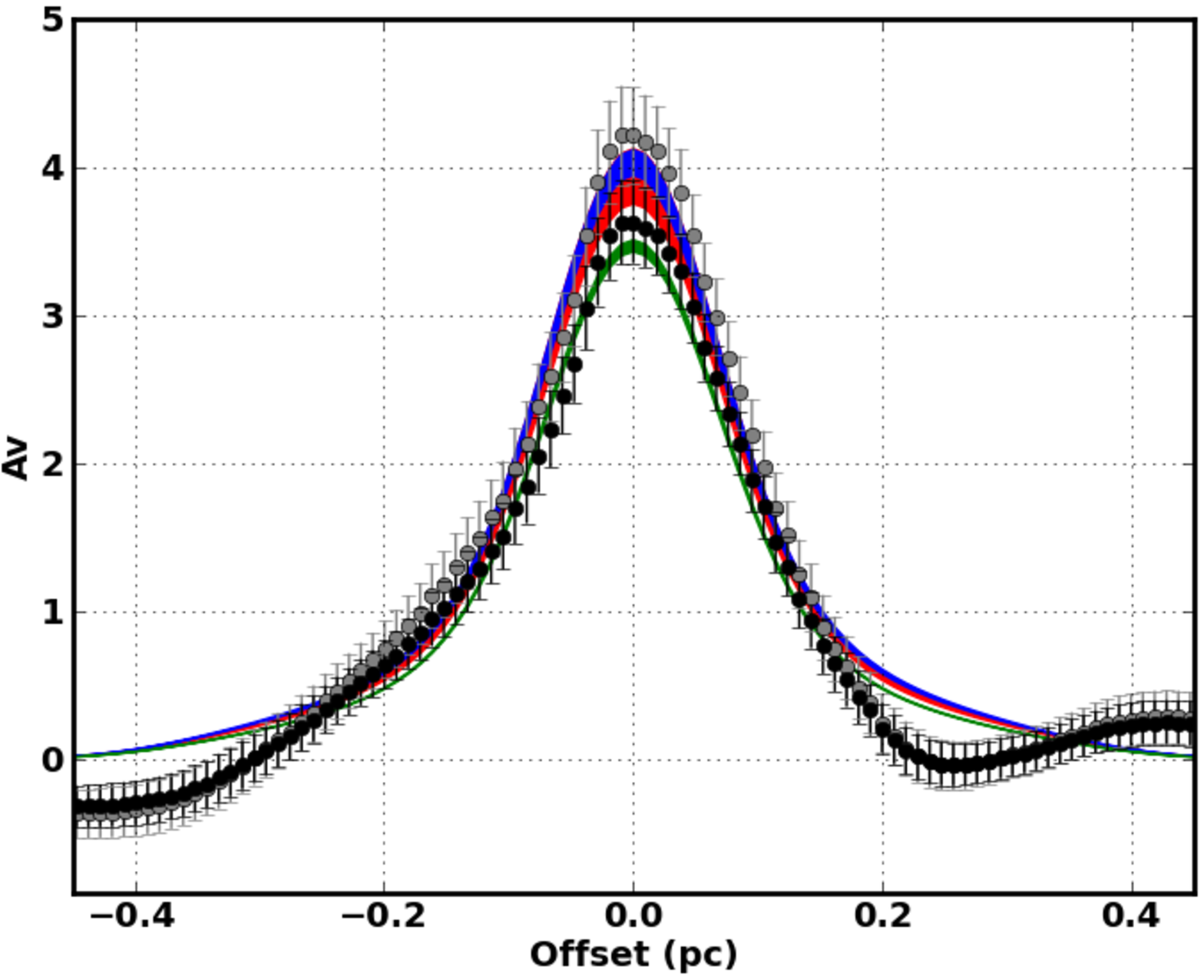}}
\caption{The grey dots trace the extinction profile of the second cut calculated from 2MASS data with $R_{{\rm V}}=3.1$ and the black ones with $R_{{\rm V}}=5.5$. The blue, red, and green areas correspond to the best-fitting models presented in Fig. \ref{filament2_emission}.}
\label{filament2_extinction} 
\end{figure}

The results for the second cut are presented in Figs. \ref{filament2_emission} and \ref{filament2_extinction}. Its emission and extinction profiles can be jointly fitted with compact aggregates for $n_T = 875 \pm 125$~H/cm$^3$, porous aggregates with 10\% of voids for $n_T = 750 \pm 250$~H/cm$^3$, or porous aggregates with 25\% of voids for $n_T = 625 \pm 125$~H/cm$^3$, which correspond to an offset of $0.12 \pm 0.02$~pc or line-of-sight with total $A_{{\rm V}} \sim 2.0-3.6$. The central and threshold densities required to model this cut are similar to what is needed for cut 1 (L1506C). 

\begin{figure}
\centerline{
\includegraphics[width=0.3\textwidth]{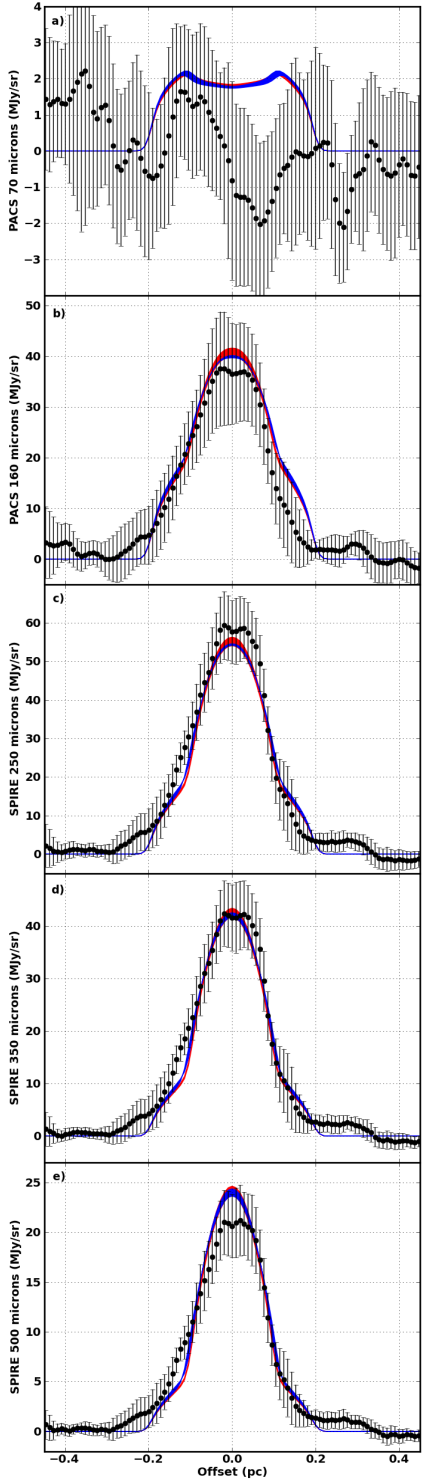}}
\caption{Brightness profiles of the fourth cut at 70, 160, 250, 350, and 500~$\mu$m (panel a, b, c, d, and e respectively). The blue area is the best-fitting models for compact aggregates and $n_T = 6\,000 \pm 500$~H/cm$^3$; the red one is for porous aggregates with 10\% of voids and $n_T = 5\,500 \pm 250$~H/cm$^3$.}
\label{filament4_emission}
\end{figure}

\begin{figure}
\centerline{
\includegraphics[width=0.3\textwidth]{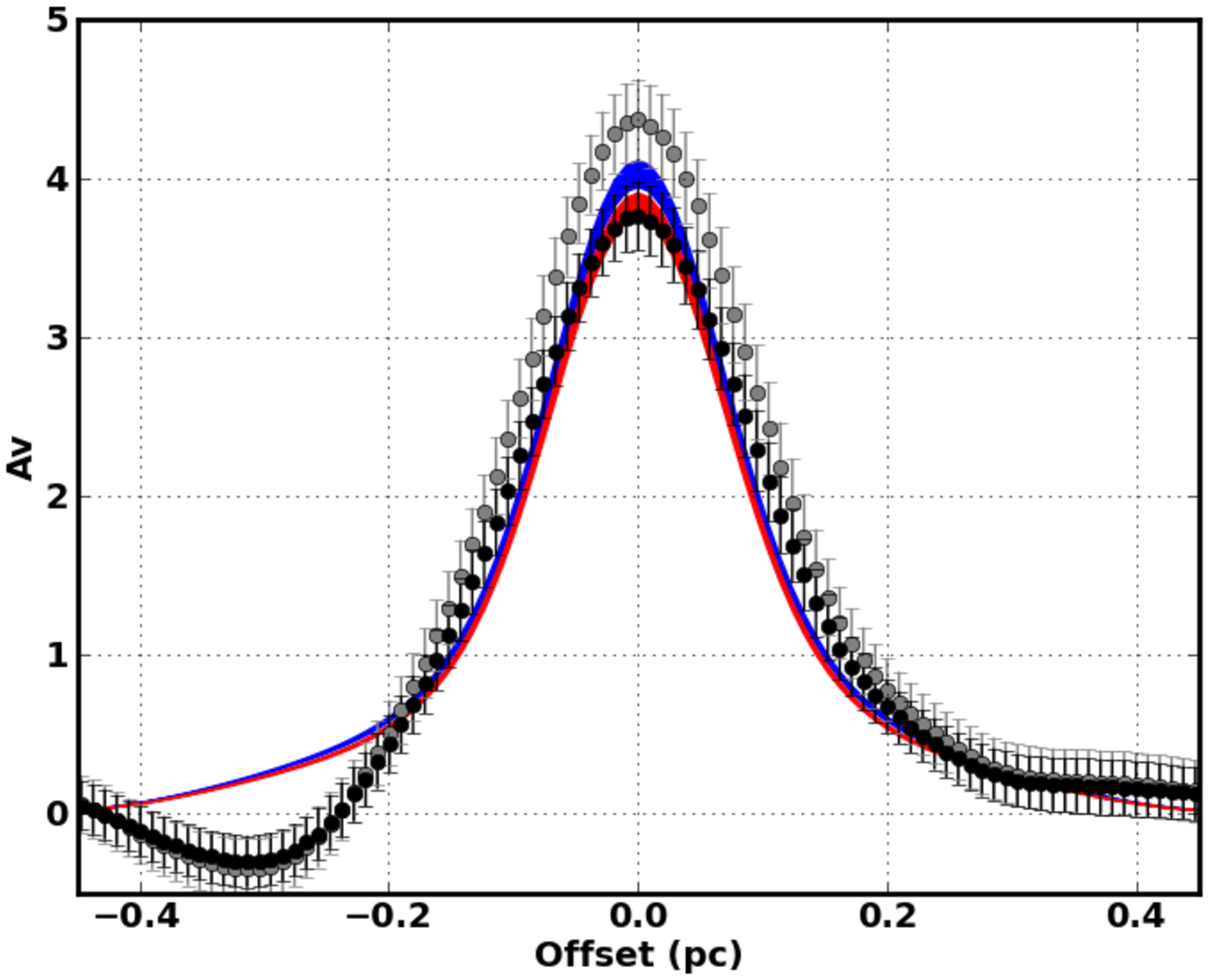}}
\caption{The grey dots trace the extinction profile of the fourth cut calculated from 2MASS data with $R_{{\rm V}}=3.1$ and the black ones with $R_{{\rm V}}=5.5$. The blue and red areas correspond to the best-fitting models presented in Fig. \ref{filament4_emission}.}
\label{filament4_extinction} 
\end{figure}

The results for the fourth cut are shown in Figs. \ref{filament4_emission} and \ref{filament4_extinction}. The best-fitting models are obtained with compact aggregates for $n_T = 6\,000 \pm 500$~H/cm$^3$ or porous aggregates with 10\% of voids for $n_T = 5\,500 \pm 250$~H/cm$^3$ in both cases. In cut 4, the central density is about ten times lower than in cuts 1 and 2 with $n_C \sim 12\,000$~H/cm$^3$, and the threshold density for the aggregates to be dominant is about 5 times higher. Thus, the aggregates seem to be present only at the very centre of the filament for offsets smaller than $0.08 \pm 0.01$~pc equivalent to total $A_{{\rm V}} \geqslant 3.1-5.2$.

For these two cuts, the lack of constraints on the central density of the gas does not allow us to infer which kind of aggregates best reproduces the data. However, we notice that for the second and the fourth cuts the aggregates with 10\% of voids, or similarly with an increase in the opacity at 250~$\mu$m by a factor of 2.16, can explain the emission and extinction profiles as for the first cut (L1506C).

\subsubsection{Third cut}
\label{section_third_cut}

Finally, we could not find any satisfactory fit for the third cut: the modelled cylinders always lead to an overestimate of the emission in the SPIRE 500~$\mu$m band by a factor of 1.3-1.4 for our best-fitting model. To check that this caveat does not come from our earlier assumption of a filament in the plane of the sky, we tested four cylinder inclinations: 10$\degr$, 20$\degr$, 30$\degr$, and 40$\degr$. The results remain unchanged with an overestimate of the emission at 500~$\mu$m. We also notice that the best-fitting models for the three other cuts predict 500~$\mu$m emission that is invariably in the upper values allowed by the error bars (Figs. \ref{filament1_emission}e, \ref{filament2_emission}e, and \ref{filament4_emission}e). This means that the optical properties of our aggregates may have an opacity spectral index that is too low in the FIR and the submm ($\beta \sim 1.3$, see Table 1 and Sect. \ref{section_caveats}). As we cannot derive physical parameters for this cut, it is excluded from the subsequent discussion.

\begin{table*}
\label{table_evolution}
\centering
\caption{Parameters of the best-fitting models for cut 1, 2, and 4, described in Sect. \ref{section_evolution}, where $R_{out}$ is the external radius of the filament and the percentages in the ``aggregates'' lines give the percentage of voids in terms of grain volume. The following lines give the factor by which the opacity at 250~$\mu$m is increased, $n_C$ is the central proton density, $R_{flat}$ is the central flat radius, $n_T$ is the threshold density above which DHGL populations disappear in favour of aggregates, $M_{line}$ is the mass per unit length of the filament, and FWHM is the width at half maximum of the hydrogen column density profile.}
\begin{tabular}{lcccccccc}
\hline
\hline
                             & Cut 1 & & \multicolumn{3}{c}{Cut 2} & & \multicolumn{2}{c}{Cut 4} \\
\hline
$R_{out}$ (pc)               & 0.30 & & \multicolumn{3}{c}{0.40}  & & \multicolumn{2}{c}{0.20}  \\
Aggregates                   & 10\%                   & & 0\% & 10\% & 25\% & & 0\% & 10\% \\
$\kappa_{250 \; \mu{\rm m}}$ & $\times$ 2.16 & & $\times$ 1.84 & $\times$ 2.16 & $\times$ 2.84 & & $\times$ 1.84 & $\times$ 2.16 \\
$n_C$ (H/cm$^3$)             & 101\,800$^{+9600}_{-3200}$ & $\;\;$ & 126\,300 $\pm$ 2\,500 & 121\,900$^{+2000}_{-6400}$ & 96\,500 $\pm$ 200 & $\;\;$ & 12\,700$^{+600}_{-700}$ & 11\,700 $\pm$ 300 \\
$R_{flat}$ (pc)                   & 0.0213                 & & 0.0152 & 0.0152 & 0.0152 & & 0.0967 & 0.0967 \\
$n_T$ (H/cm$^3$)             & 1\,500 $\pm$ 250 & & 875 $\pm$ 125 & 750 $\pm$ 250 & 625 $\pm$ 125 & & 6\,000 $\pm$ 500 & 5\,500 $\pm$ 250 \\
$M_{line}$ (M$_{\odot}$/pc)  & 18.3$^{+1.7}_{-0.8}$ & & 14.7 $\pm$ 0.3 & 14.2$^{+0.2}_{-0.7}$ & 11.2 $\pm$ 0.02 & & 15.3$^{+0.8}_{-0.9}$ & 14.1 $\pm$ 0.4 \\
FWHM (pc) & $0.069 \pm 0.005$ & & \multicolumn{3}{c}{$0.051 \pm 0.003$} & & \multicolumn{2}{c}{$0.23 \pm 0.01$} \\
\hline
\end{tabular}
\end{table*}

\subsubsection{Temperature distribution}
\label{section_temperature_distribution}

\begin{figure}
\centerline{
\includegraphics[width=0.35\textwidth]{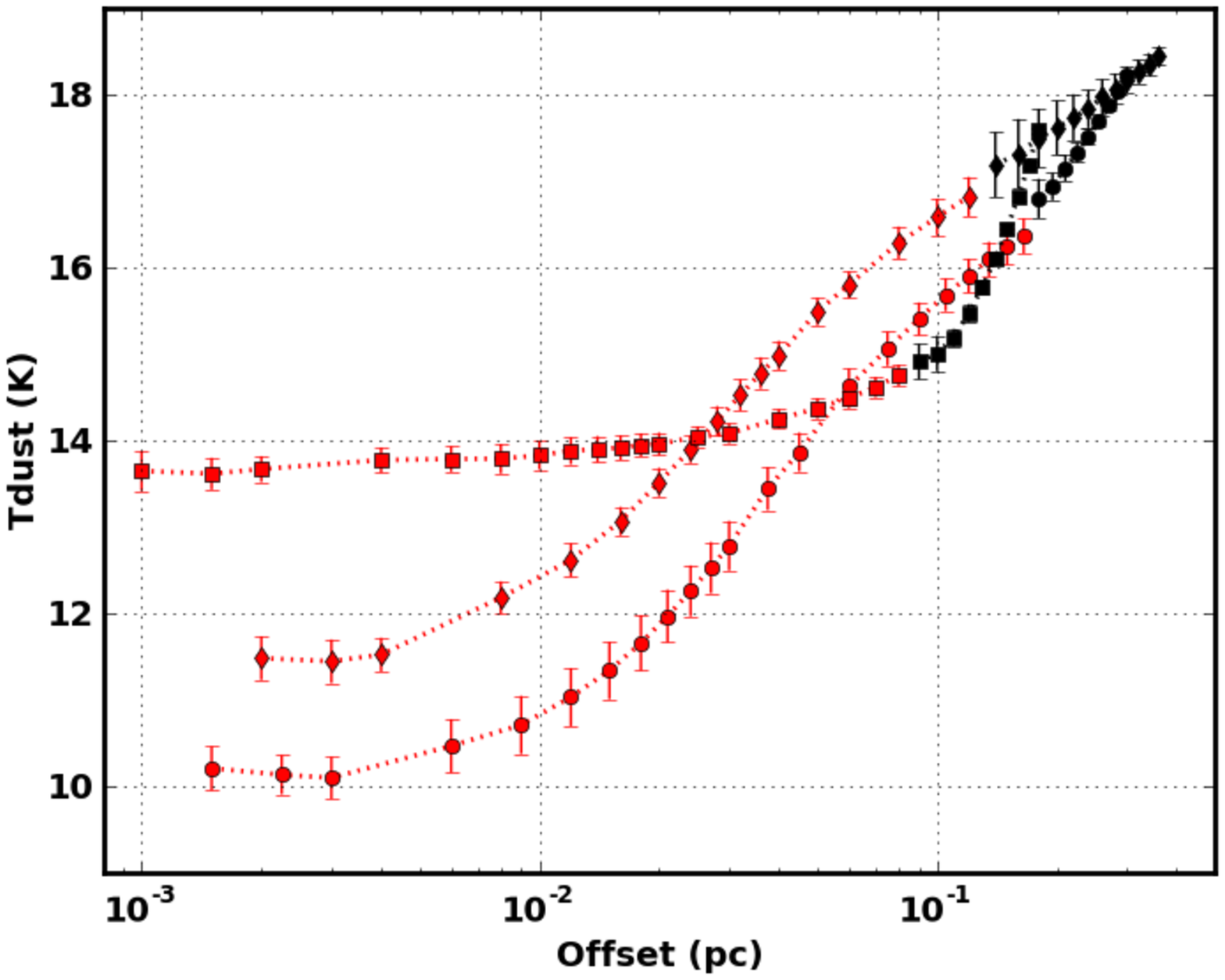}}
\caption{Temperature distributions of the best-fitting models for the first (bullets), the second (diamonds), and the fourth cut (squares). The red symbols are the equilibrium temperature of the aggregates with 10\% of voids, averaged over their size distribution, and the black symbols are the same for the mixture of large grains of the DHGL populations (LamC and aSil).}
\label{fig_Tdust} 
\end{figure}

Figure \ref{fig_Tdust} presents the temperature distributions that correspond to the best-fitting models obtained for aggregates with 10\% of voids ($\kappa_{250~\mu{\rm m}} \times 2.16$) as a function of the radius for cuts 1, 2, and 4. These temperatures are the equilibrium temperatures of the grains, averaged over the size distribution. The central temperatures of the second and the fourth cuts are $T_{dust} = 11.3 \pm 0.3$~K and $T_{dust} = 13.6 \pm 0.2$~K, respectively. For the first cut, we find it is significantly lower with $T_{dust} = 10.2 \pm 0.3$~K. The fact that this value is very close to the gas temperature of about 10~K used by \citet{Pagani2010} at the centre of L1506C is probably coincidental. Indeed for gas densities lower than $10^5$~H/cm$^3$, a mild dust-gas coupling is expected. In this case, the dust grains are usually colder than the gas by 2-3~K, especially when the grain growth is taken into account because it further decreases the dust-gas coupling \citep{Zhilkin2009, Juvela2011, Nielbock2012}. This indicates that the gas temperature in \citet{Pagani2010} may have been slightly underestimated.

\subsection{Column density distribution}
\label{filament_width}

\begin{figure}
\centerline{
\includegraphics[width=0.35\textwidth]{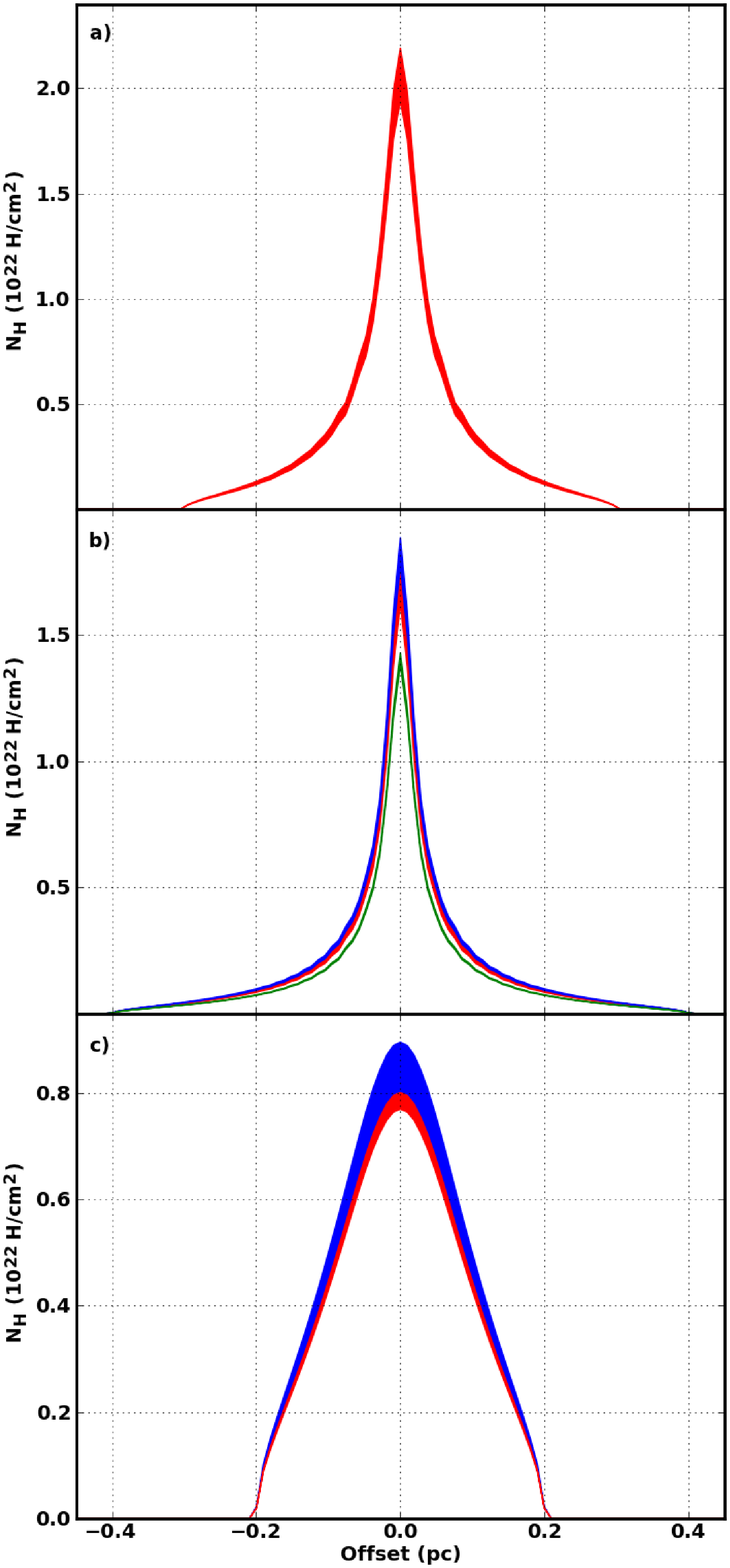}}
\caption{Column density distributions of the best-fitting models for the cuts 1 (a), 2 (b), and 4 (c), as a function of the offset from the filament centre. The blue areas show the best-fitting models using compact aggregates, the red areas using aggregates with 10\% of voids, and the green areas with 25\% of voids.}
\label{fig_filament_width} 
\end{figure}

Figure \ref{fig_filament_width} shows the hydrogen column density distributions corresponding to the best-fitting models for cuts 1, 2, and 4 (see Sect. \ref{section_first_cut} and \ref{section_second_fourth_cuts}). Their full widths at half maximum are FWHM = $0.069 \pm 0.005$, $0.051 \pm 0.003$, and $0.23 \pm 0.01$~pc for cuts 1, 2, and 4, reaching $N_H \sim 1.9 \times 10^{22}$, $1.5 \times 10^{22}$, and $0.8 \times 10^{22}$~H/cm$^2$, respectively. The width of L1506 is of the same order of that found by \citet{Arzoumanian2011}, who studied a large sample of 90 filaments in the IC5146, Polaris, and Aquila regions. Averaging along the length of the filaments, they found an almost constant width of $0.10 \pm 0.03$~pc. However, our analysis shows that the width is not constant along the filament and that it varies by a factor $\sim 4$ according to the local column density in agreement with \citet{Juvela2012}. We find that the smaller widths correspond to the densest parts of L1506, which agrees with the idea of a gravitationally contracting cloud.

An anti-correlation between the FWHM and the central column density was found by \citet{Fischera2012}, who analysed analytically isothermal cylinders confined by an external pressure. The FWHM that we find for cuts 1 and 2 match their values for a self-gravitating cloud (they computed FWHM $\sim 0.05$~pc for $N_H \sim 2 \times 10^{22}$~H/cm$^2$, see their Fig. 8). However, we find that the FWHM of the fourth cut is much higher than any of their analytical values. The fourth cut corresponds to the less dense area that we studied. It has a central density 10 times lower than in the other cuts and a central column density approximately twice as low. \citet{Heitsch2013} modelled turbulent isothermal hydrostatic infinite cylinders and explored the effect of gas accretion on the cloud structures. They showed that at low column densities the FWHM and the central $N_H$ are expected to be uncorrelated with FWHM = 0.2-0.3~pc, which is similar to the width of the fourth cut. For $N_H > 10^{22}$~H/cm$^2$, their calculations show a decreasing FWHM with increasing $N_H$, agreeing with the width of the first and second cuts.

\section{Discussion}
\label{section_discussion}

We investigate here the possibility for the aggregates to form inside the dense filament in terms of coagulation timescales using the parameters of the best-fitting models, as described in the previous section for the cuts 1, 2, and 4. We also consider the stability of the filament against gravitational collapse.

\subsection{Coagulation timescale}
\label{coagulation_timescale}

\zero
\begin{table}
\label{table_size_density}
\centering
\caption{Averaged sizes in nanometres,$\langle a \rangle$, and densities, $n_{dust}/n_H$, for the DHGL populations. The four following lines of the table give the critical speed between the species in m/s (see Eq. \ref{equation_vcrit}). The last line gives $v_{crit}$ for collisions with an aggregate containing 10\% of voids ($\langle a \rangle \sim 20$~nm).}
\begin{tabular}{lcccc}
\hline
\hline
Dust type      & PAH                  & SamC                 & LamC                  & aSil \\
\hline
$\langle a \rangle$\tablefootmark{a} & 0.6        & 2                    & 9                     & 7   \\
$n_{dust}/n_H$ & $5.1 \times 10^{-7}$ & $2.6 \times 10^{-9}$ & $2.0 \times 10^{-11}$ & $2.4 \times 10^{-10}$ \\
PAH            & 804                  & 563                  & 478                   & 485 \\
SamC           & 563                  & 299                  & 200                   & 209 \\
LamC           & 478                  & 200                  & 89                    & 99  \\
aSil           & 485                  & 209                  & 99                    & 110 \\
10\%           & 463                  & 182                  & 68                    & 79  \\
\hline
\end{tabular}
\tablefoot{\tablefoottext{a}{These sizes correspond to the average size over the size distribution in terms of grain number. In terms of mass, the average sizes are of 0.7 and 3.5~nm for the PAHs and the SamC, respectively, the size distributions of which are log-normals. For the LamC and the aSil grains, for which the size distributions decrease as power-laws, they are 221 and 255~nm, respectively.}}
\end{table}

\begin{table}
\label{table_caogulation_time}
\centering
\caption{Coagulation timescales in years for cuts 1, 2, and 4. These are the coagulation timescales for the sticking of only two grains with a relative speed equal to the turbulent speed, $\Delta v = 68$~m/s.}
\begin{tabular}{lccc}
\hline
\hline
             & $\tau_1$ (yrs)    & $\tau_2$ (yrs)    & $\tau_4$ (yrs)    \\
\hline
\hline
PAH on LamC  & 70                & 60                & 580               \\
PAH on aSil  & 80                & 60                & 660               \\
SamC on LamC & $8.5 \times 10^3$ & $7.1 \times 10^3$ & $7.3 \times 10^4$ \\
SamC on aSil & $9.6 \times 10^3$ & $8.0 \times 10^3$ & $8.3 \times 10^4$ \\
LamC on LamC & $4.4 \times 10^5$ & $3.7 \times 10^5$ & $3.8 \times 10^6$ \\
LamC on aSil & $4.9 \times 10^5$ & $4.1 \times 10^5$ & $4.3 \times 10^6$ \\
aSil on LamC & $4.0 \times 10^4$ & $3.4 \times 10^4$ & $3.5 \times 10^5$ \\
aSil on aSil & $4.5 \times 10^4$ & $3.7 \times 10^4$ & $3.9 \times 10^5$ \\
\hline
\end{tabular}
\end{table}

As seen in Sect. \ref{section_modelling}, the dust optical properties must vary from the outer layers of the filament to its denser centre to explain its emission and extinction profiles. Following some rough assumptions, we now estimate whether the relatively low density of this filament ($n_H \lesssim 10^5$~H/cm$^3$) allows for the formation of aggregates on a reasonable timescale. To answer this question, we use the simple approach described below.

The coagulation timescale depends on the grain densities, their sizes through the collisional cross-section, and the critical speed above which a collision would lead to destruction instead of aggregation. The average sizes and the densities of the four DHGL populations are listed in Table 4. According to \citet{Dominik1997} and \citet{Stepnik2003}, the critical speed for a destructive collision between two species 1 and 2 of sizes $a_1$ and $a_2$ can be expressed as
\begin{equation}
\label{equation_vcrit}
v_{crit} = 670 \left[ \left( \frac{1}{a_1} + \frac{1}{a_2} \right) 10^{-5} \; {\rm cm} \right]^{5/6} \; {\rm (cm/s).}
\end{equation}
The critical speeds for collisions between species of the DHGL populations are given in Table 4. Then, the coagulation timescale of two species depends on their cross-sections and densities, and on their relative velocity, $\Delta v$. Under most molecular cloud conditions, $\Delta v$ is defined by the turbulence \citep{Ossenkopf1993, Ormel2007, Ormel2009}. Because of its own inertia, a grain does not instantaneously follow the gas motion and as a result, the maximum velocity difference between two grains is the gas turbulent velocity. This means that in any case $\Delta v \leqslant v_{turb}$, where the actual velocity difference depends on the efficiency of the gas-dust coupling. In L1506C, the modelling of the C$^{18}$O and N$_2$H$^+$ lines at the centre of the cloud demonstrated a low turbulent velocity, $v_{turb} \leqslant 68$~m/s \citep{Pagani2010}. This is lower than $v_{crit}$ for most of the possible collisions among DHGL populations. As an upper limit, we assume that $\Delta v = v_{turb}$, which leads to the shortest possible coagulation timescale, $\tau_{coag}$. Following \citet{Stepnik2003}, we define the coagulation timescale of one particle of species 1 onto one particle of species 2 as
\begin{equation}
\tau_{coag} = \left[ \pi (a_1 + a_2)^2 n_1 \Delta v\right]^{-1},
\end{equation}
where $n_1$ is the density of the species 1. We define the average $\tau_{coag}$ between two species with size distributions $n(a_1)$ and $n(a_2)$, respectively, as
\begin{equation}
\tau_{coag}^{-1} = \int_{a_{1}^{min}}^{a_{1}^{max}} n(a_1) da_1 \int_{a_{2}^{min}}^{a_{2}^{max}} n(a_2) da_2 \; \pi (a_1+a_2)^2 n_1 \Delta v.
\end{equation}
This relation supposes that all the collisions lead to coagulation, which again puts a lower limit on $\tau_{coag}$. The corresponding values are given in Table 5 for cut 1, 2, and 4.

However, the aggregates required to explain the emission and extinction profiles of L1506 contain more than two grains: $\langle a \rangle \sim 0.02$~$\mu$m for the aggregates with 10\% of voids. According to \citet{Koehler2012}, the coagulation of at least four big grains is required to increase the opacity at 250~$\mu$m by a factor of 2-2.5. We assume that the average aggregate is made of three aSil grains plus one LamC grain. As a first approximation, the coagulation timescale to stick three aSil grains together is supposed to be about twice the coagulation time of one aSil grain onto another aSil grain. Then, a LamC grain can stick onto this aSil aggregate or the aSil aggregate can stick onto a LamC grain. This leads to coagulation timescales of $\tau_{coag} \sim 1.3-2.8 \times 10^5$~yrs for cut 1, $\tau_{coag} \sim 1.1-2.3 \times 10^5$~yrs for cut 2, and $\tau_{coag} \sim 1.1-2.4 \times 10^6$~yrs for cut 4. The inclusion of small grains (PAHs and SamC) in the aggregates is very fast, so it can be neglected in our rough estimate of $\tau_{coag}$.

Now that we have an estimate of $\tau_{coag}$, it can be compared to the cloud lifetime. This lifetime is at least equal to the free-fall timescale. According to \citet{Kawachi1998}, the free-fall timescale of an isothermal cylinder with a constant density $n_0$ is
\begin{equation}
\tau_{ff} = \frac{1}{\sqrt{4 \pi G m_H n_0}},
\end{equation}
where G is the gravitational constant and the $m_H$ the mass of the hydrogen atom. Considering the averaged density of our best-fitting models, $\langle n \rangle_1 \sim 11\,000$, $\langle n \rangle_2 \sim 7\,000$, and $\langle n \rangle_4 \sim 6\,500$~H/cm$^3$, we get $\tau_{coag} \sim 2.6 \times 10^5$, $3.2 \times 10^5$, and $3.3 \times 10^5$~yrs for the cuts 1, 2, and 4, respectively.
Thus, the coagulation and free-fall timescales are of the same order. Because these free-fall timescales are lower limits for the cloud lifetime since we neglected any support, which would be able to slow down the gravitational collapse (e.g., the magnetic field or the rotation of the filament around its axis of symmetry), our results support the idea of a grain coagulation taking place inside the filament.

\subsection{Stability of the filament}
\label{section_stability}

\citet{Pagani2010} showed that the velocity profiles of C$^{18}$O, and N$_2$H$^+$ prove an inward motion of the gas of at least 0.1~km/s in L1506C. They interpreted this as a proof of the collapse of the core. We can further test this finding by computing the masses per unit length of our best-fitting models:
\begin{eqnarray}
M_{line} &=& \int_{0}^{R_{out}} 2 \pi n_H(r) r dr \\
M_{line} &=& \pi n_C R_{flat}^2 \ln\left[ 1 + \left( \frac{R_{out}}{R_{flat}} \right)^2 \right].
\end{eqnarray}
This leads to $M_{line} = 18.3^{+1.7}_{-0.8}$, $14.2^{+0.2}_{-0.7}$, and $14.1 \pm 0.4$~M$_{\odot}$/pc for the cuts 1, 2, and 4, respectively. These linear masses can be compared to the critical mass defined by \citet{Ostriker1964}, above which the cloud collapses: $M_{crit} = 2c_S^2 / G$, where $c_S$ is the sound speed and depends on the gas temperature. Following the prescription of \citet{Pagani2010}, we assume a gas temperature of 10~K, which gives $M_{crit} = 16.8$~M$_{\odot}$/pc. Our results agree with the results of \citet{Pagani2010}, who found that the filament may be collapsing at the location of the first cut or L1506C ($M_{line} \gtrsim M_{crit}$). At the two other locations along L1506, the linear masses are slightly lower. As we lack information on the gas temperature for cuts 2 and 4 and as the equilibrium state of a cloud depends on many parameters (external pressure, turbulence, and magnetic field), we cannot conclude on the stability of the filament at these locations.

\subsection{Main results and caveats}
\label{section_caveats}

With the advantage of the improved sensitivity, resolution, and spectral coverage from the Herschel PACS and SPIRE instruments, we studied the variations in the dust properties between the diffuse ISM and the centre of the L1506 cloud. To do so, we tested the data against the dust models described in \citet{Compiegne2011} and \citet{Ysard2012}. Our subsequent findings are the following:
\begin{itemize}
\item[\it (i)] the emission and extinction profiles of L1506 cannot be explained with dust grains similar to those found in the diffuse ISM;
\item[\it (ii)] they cannot be explained by a model where we consider that all the grains have coagulated to form compact or fluffy aggregates;
\item[\it (iii)] they can be reproduced by considering that the grains are of diffuse ISM type in the outer layers of the filament, whereas aggregates dominate the dust abundance above a certain density threshold (typically a few 10$^3$~H/cm$^3$).
\end{itemize}

Considering the first point, we tested the data against two different diffuse dust models \citep{Draine2007, Compiegne2011} and for several filament inclination angles. The results were unchanged. It was impossible to fit the emission profiles and the extinction was systematically overestimated. This also matches many previous studies that used various techniques to reach this result (see references in Sect. \ref{section_introduction}). Thus, we consider it is a definitive result that the emission and extinction in L1506 cannot be explained with grains of the diffuse ISM type.

At first, the second point appears more questionable. Indeed, we find that the far-IR emission can be easily explained with a single aggregate population, whereas the extinction is underestimated (for all the inclination angles tested, $0\degr \leqslant \alpha \leqslant 40\degr$). This result may come from our aggregate model. The question is: can the data be explained by an aggregate model with a different near-IR to submm opacity ratio? In Fig. \ref{compare_kappa}, we show three different ``evolved grain'' models. The opacity of the dust model of \citet{Weingartner2001} at 250~$\mu$m does not seem appropriate in explaining the data, as it is of the order of the opacity of the dust in the diffuse ISM. The aggregate models of \citet{Ormel2011} and \citet{Ossenkopf1994} have near- and far-IR opacities that are similar to those of our model \citep{Ysard2012}: their far-IR opacity differs from ours by $\sim 30$\% and in the visible and near-IR by $\sim 15$\%. This is far from reaching the factors of 1.5 to 2.5 reported in Sect. \ref{section_aggregates} and by which our best-fitting models underestimate the extinction. These models may not represent the entire plausible range of dust models; however, it is an indication of the robustness of our result. Detailed studies for more molecular clouds, which undergo various environmental conditions, are required to strengthen the fact that no single dust model can simultaneously fit the dust emission and extinction (i.e., a variation in dust properties along the line-of-sight is required to simultaneously fit both the emission and the extinction).

If one trusts the robustness of the previous result, the third point comes naturally: diffuse ISM-type grains are found at the edge of the cloud and aggregates at the centre. This result is strengthened by the correspondence between our results and the modelling of the molecular gas lines performed by \citet{Pagani2010}. However, the exact numerical values (e.g., Table 3), especially the threshold density above which the aggregates dominate the dust abundance, are model-dependent. For instance, the simple addition of an ice mantle on top of the aggregates may lead to significantly different near-IR to far-IR opacity ratios, as can be seen in Fig. \ref{compare_kappa}.

This study made us aware of an intrinsic caveat of our aggregate model. As shown in Sect. \ref{section_third_cut}, our aggregate opacity spectral index is too low in the far-IR and the submillimetre with $\beta \sim 1.3$. Because of this low value, we are not able to find any satisfactory fit for the 500~$\mu$m emission profile of the third cut, and the best-fitting models for the three other cuts always lead to higher estimates of this same profile. This low $\beta$-value is due to the wavelength dependence of the absorption efficiency ($Q_{abs}$) of the carbon grains present in the DHGL model. Performing Mie calculations, \citet{Compiegne2011} defined their absorption and extinction efficiencies from the BE sample complex refractive index data derived from laboratory measurements by \citet{Zubko1996}. This sample was assumed to be a good analogous to interstellar (hydrogenated) amorphous carbon. However, it appears that this experimental sample might not be the most appropriate to account for the interstellar carbonaceous dust.

\section{Conclusions}
\label{section_conclusions}

We performed detailed modelling of the dust content and density structure of L1506, a dense filament in the Taurus molecular complex, and included full-radiative transfer calculations. We definitively excluded the possibility that the emission and extinction profiles of this filament could be explained using the properties of the dust found in the standard diffuse ISM. We showed that the dust far-IR opacity has to increase from the outer to the inner parts of L1506. We interpret this increase in the opacity as the formation of large aggregates. To fit the far-IR dust emission and extinction profiles simultaneously along the filament, aggregates with an average size in terms of mass of about 0.4~$\mu$m and an increase in the opacity at 250~$\mu$m of 1.8 to 2.2 are required. They have to prevail over diffuse ISM-type grains, when the local gas density reaches a few 1\,000~H/cm$^3$. The exact numerical value of this threshold naturally depends on the dust model chosen to fit the data. The size of these aggregates is not far from what is required to explain the coreshine observed in molecular clouds \citep{Pagani_core2010, Steinacker2010}. Using a simple approach, we showed that these aggregates may have time to form inside the filament within the cloud lifetime. Our best-fitting models finally reveal that the width of the filament varies according to the column density. More modelling of dense filaments is needed to confirm our results. An important breakthrough may thus arise from detailed modelling including both the emitted and scattered light in addition to the extinction.

\acknowledgements{N.Y. acknowledges the support of a CNES post-doctoral research grant. We thank our anonymous referee, whose useful comments helped us to improve the content of this paper. SPIRE has been developed by a consortium of institutes led by Cardiff Univ. (UK) and including: Univ. Lethbridge (Canada);
NAOC (China); CEA, LAM (France); IFSI, Univ. Padua (Italy); IAC (Spain); Stockholm Observatory (Sweden); Imperial College London, RAL, UCL-MSSL, UKATC, Univ. Sussex (UK); and Caltech, JPL, NHSC, Univ. Colorado (USA). This development has been supported by national funding agencies: CSA (Canada); NAOC (China); CEA, CNES, CNRS (France); ASI (Italy); MCINN (Spain); SNSB (Sweden); STFC, UKSA (UK); and NASA (USA).}

\bibliography{biblio}

\begin{thebibliography}{73}
\expandafter\ifx\csname natexlab\endcsname\relax\def\natexlab#1{#1}\fi

\bibitem[{{Abergel} {et~al.}(2010){Abergel}, {Arab}, {Compi{\`e}gne}, {Kirk},
  {Ade}, {Anderson}, {Andr{\'e}}, {Baluteau}, {Bernard}, {Blagrave},
  {Bontemps}, {Boulanger}, {Cohen}, {Cox}, {Dartois}, {Davis}, {Emery},
  {Fulton}, {Gry}, {Habart}, {Huang}, {Joblin}, {Jones}, {Lagache}, {Lim},
  {Madden}, {Makiwa}, {Martin}, {Miville-Desch{\^e}nes}, {Molinari}, {Moseley},
  {Motte}, {Naylor}, {Okumura}, {Pinheiro Gon{\c c}alves}, {Polehampton},
  {Rodon}, {Russeil}, {Saraceno}, {Sauvage}, {Sidher}, {Spencer}, {Swinyard},
  {Ward-Thompson}, {White}, \& {Zavagno}}]{Abergel2010}
{Abergel}, A., {Arab}, H., {Compi{\`e}gne}, M., {et~al.} 2010, \aap, 518, L96

\bibitem[{{Abergel} {et~al.}(1996){Abergel}, {Bernard}, {Boulanger},
  {Cesarsky}, {Desert}, {Falgarone}, {Lagache}, {Perault}, {Puget}, {Reach},
  {Nordh}, {Olofsson}, {Huldtgren}, {Kaas}, {Andre}, {Bontemps}, {Burgdorf},
  {Copet}, {Davies}, {Montmerle}, {Persi}, \& {Sibille}}]{Abergel1996}
{Abergel}, A., {Bernard}, J.~P., {Boulanger}, F., {et~al.} 1996, \aap, 315,
  L329

\bibitem[{{Abergel} {et~al.}(1994){Abergel}, {Boulanger}, {Mizuno}, \&
  {Fukui}}]{Abergel1994}
{Abergel}, A., {Boulanger}, F., {Mizuno}, A., \& {Fukui}, Y. 1994, \apjl, 423,
  L59

\bibitem[{{Andr{\'e}} {et~al.}(2010){Andr{\'e}}, {Men'shchikov}, {Bontemps},
  {K{\"o}nyves}, {Motte}, {Schneider}, {Didelon}, {Minier}, {Saraceno},
  {Ward-Thompson}, {di Francesco}, {White}, {Molinari}, {Testi}, {Abergel},
  {Griffin}, {Henning}, {Royer}, {Mer{\'{\i}}n}, {Vavrek}, {Attard},
  {Arzoumanian}, {Wilson}, {Ade}, {Aussel}, {Baluteau}, {Benedettini},
  {Bernard}, {Blommaert}, {Cambr{\'e}sy}, {Cox}, {di Giorgio}, {Hargrave},
  {Hennemann}, {Huang}, {Kirk}, {Krause}, {Launhardt}, {Leeks}, {Le Pennec},
  {Li}, {Martin}, {Maury}, {Olofsson}, {Omont}, {Peretto}, {Pezzuto}, {Prusti},
  {Roussel}, {Russeil}, {Sauvage}, {Sibthorpe}, {Sicilia-Aguilar}, {Spinoglio},
  {Waelkens}, {Woodcraft}, \& {Zavagno}}]{Andre2010}
{Andr{\'e}}, P., {Men'shchikov}, A., {Bontemps}, S., {et~al.} 2010, \aap, 518,
  L102

\bibitem[{{Arzoumanian} {et~al.}(2011){Arzoumanian}, {Andr{\'e}}, {Didelon},
  {K{\"o}nyves}, {Schneider}, {Men'shchikov}, {Sousbie}, {Zavagno}, {Bontemps},
  {di Francesco}, {Griffin}, {Hennemann}, {Hill}, {Kirk}, {Martin}, {Minier},
  {Molinari}, {Motte}, {Peretto}, {Pezzuto}, {Spinoglio}, {Ward-Thompson},
  {White}, \& {Wilson}}]{Arzoumanian2011}
{Arzoumanian}, D., {Andr{\'e}}, P., {Didelon}, P., {et~al.} 2011, \aap, 529, L6

\bibitem[{{Bazell} \& {Dwek}(1990)}]{Bazell1990}
{Bazell}, D. \& {Dwek}, E. 1990, \apj, 360, 142

\bibitem[{{Bergin} \& {Tafalla}(2007)}]{Bergin2007}
{Bergin}, E.~A. \& {Tafalla}, M. 2007, \araa, 45, 339

\bibitem[{{Bernard} {et~al.}(1999){Bernard}, {Abergel}, {Ristorcelli}, {Pajot},
  {Torre}, {Boulanger}, {Giard}, {Lagache}, {Serra}, {Lamarre}, {Puget},
  {Lepeintre}, \& {Cambr{\'e}sy}}]{Bernard1999}
{Bernard}, J.~P., {Abergel}, A., {Ristorcelli}, I., {et~al.} 1999, \aap, 347,
  640

\bibitem[{Bohren \& Huffman(1983)}]{Bohren1983}
Bohren, C.~F. \& Huffman, D.~R. 1983, New York: Wiley

\bibitem[{{Cambr{\'e}sy} {et~al.}(2001){Cambr{\'e}sy}, {Boulanger}, {Lagache},
  \& {Stepnik}}]{Cambresy2001}
{Cambr{\'e}sy}, L., {Boulanger}, F., {Lagache}, G., \& {Stepnik}, B. 2001,
  \aap, 375, 999

\bibitem[{{Cardelli} {et~al.}(1989){Cardelli}, {Clayton}, \&
  {Mathis}}]{Cardelli1989}
{Cardelli}, J.~A., {Clayton}, G.~C., \& {Mathis}, J.~S. 1989, \apj, 345, 245

\bibitem[{{Compi{\`e}gne} {et~al.}(2011){Compi{\`e}gne}, {Verstraete}, {Jones},
  {Bernard}, {Boulanger}, {Flagey}, {Le Bourlot}, {Paradis}, \&
  {Ysard}}]{Compiegne2011}
{Compi{\`e}gne}, M., {Verstraete}, L., {Jones}, A., {et~al.} 2011, \aap, 525,
  A103

\bibitem[{{Dominik} \& {Tielens}(1997)}]{Dominik1997}
{Dominik}, C. \& {Tielens}, A.~G.~G.~M. 1997, \apj, 480, 647

\bibitem[{{Dorschner} \& {Henning}(1995)}]{Dorschner1995}
{Dorschner}, J. \& {Henning}, T. 1995, \aapr, 6, 271

\bibitem[{{Draine}(2003)}]{Draine2003}
{Draine}, B.~T. 2003, \araa, 41, 241

\bibitem[{{Draine} \& {Anderson}(1985)}]{Draine1985}
{Draine}, B.~T. \& {Anderson}, N. 1985, \apj, 292, 494

\bibitem[{{Draine} \& {Li}(2007)}]{Draine2007}
{Draine}, B.~T. \& {Li}, A. 2007, \apj, 657, 810

\bibitem[{{Fischera} \& {Martin}(2012{\natexlab{a}})}]{Fischera2012}
{Fischera}, J. \& {Martin}, P.~G. 2012{\natexlab{a}}, \aap, 547, A86

\bibitem[{{Fischera} \& {Martin}(2012{\natexlab{b}})}]{Fischera2012b}
{Fischera}, J. \& {Martin}, P.~G. 2012{\natexlab{b}}, \aap, 542, A77

\bibitem[{{Fitzpatrick}(1999)}]{Fitzpatrick1999}
{Fitzpatrick}, E.~L. 1999, \pasp, 111, 63

\bibitem[{{Flagey} {et~al.}(2009){Flagey}, {Noriega-Crespo}, {Boulanger},
  {Carey}, {Brooke}, {Falgarone}, {Huard}, {McCabe}, {Miville-Desch{\^e}nes},
  {Padgett}, {Paladini}, \& {Rebull}}]{Flagey2009}
{Flagey}, N., {Noriega-Crespo}, A., {Boulanger}, F., {et~al.} 2009, \apj, 701,
  1450

\bibitem[{{Goldsmith} {et~al.}(2008){Goldsmith}, {Heyer}, {Narayanan}, {Snell},
  {Li}, \& {Brunt}}]{Goldsmith2008}
{Goldsmith}, P.~F., {Heyer}, M., {Narayanan}, G., {et~al.} 2008, \apj, 680, 428

\bibitem[{{Griffin} {et~al.}(2010){Griffin}, {Abergel}, {Abreu}, {Ade},
  {Andr{\'e}}, {Augueres}, {Babbedge}, {Bae}, {Baillie}, {Baluteau}, {Barlow},
  {Bendo}, {Benielli}, {Bock}, {Bonhomme}, {Brisbin}, {Brockley-Blatt},
  {Caldwell}, {Cara}, {Castro-Rodriguez}, \& {Cerulli}}]{Griffin2010}
{Griffin}, M.~J., {Abergel}, A., {Abreu}, A., {et~al.} 2010, \aap, 518, L3

\bibitem[{{Hartmann}(2002)}]{Hartmann2002}
{Hartmann}, L. 2002, \apj, 578, 914

\bibitem[{{Heitsch}(2013)}]{Heitsch2013}
{Heitsch}, F. 2013, \apj, 769, 115

\bibitem[{{Hill} {et~al.}(2011){Hill}, {Motte}, {Didelon}, {Bontemps},
  {Minier}, {Hennemann}, {Schneider}, {Andr{\'e}}, {Men`Shchikov}, {Anderson},
  {Arzoumanian}, {Bernard}, {di Francesco}, {Elia}, {Giannini}, {Griffin},
  {K{\"o}nyves}, {Kirk}, {Marston}, {Martin}, {Molinari}, {Nguyn Lu'O'Ng},
  {Peretto}, {Pezzuto}, {Roussel}, {Sauvage}, {Sousbie}, {Testi},
  {Ward-Thompson}, {White}, {Wilson}, \& {Zavagno}}]{Hill2011}
{Hill}, T., {Motte}, F., {Didelon}, P., {et~al.} 2011, \aap, 533, A94+

\bibitem[{{Juvela}(2005)}]{Juvela2005}
{Juvela}, M. 2005, \aap, 440, 531

\bibitem[{{Juvela} {et~al.}(2008){Juvela}, {Pelkonen}, {Padoan}, \&
  {Mattila}}]{Juvela2008}
{Juvela}, M., {Pelkonen}, V.-M., {Padoan}, P., \& {Mattila}, K. 2008, \aap,
  480, 445

\bibitem[{{Juvela} {et~al.}(2012){Juvela}, {Ristorcelli}, {Pagani}, {Doi},
  {Pelkonen}, {Marshall}, {Bernard}, {Falgarone}, {Malinen}, {Marton},
  {McGehee}, {Montier}, {Motte}, {Paladini}, {T{\'o}th}, {Ysard}, {Zahorecz},
  \& {Zavagno}}]{Juvela2012}
{Juvela}, M., {Ristorcelli}, I., {Pagani}, L., {et~al.} 2012, \aap, 541, A12

\bibitem[{{Juvela} {et~al.}(2011){Juvela}, {Ristorcelli}, {Pelkonen},
  {Marshall}, {Montier}, {Bernard}, {Paladini}, {Lunttila}, {Abergel},
  {Andr{\'e}}, {Dickinson}, {Dupac}, {Malinen}, {Martin}, {McGehee}, {Pagani},
  {Ysard}, \& {Zavagno}}]{JuvelaCC2011}
{Juvela}, M., {Ristorcelli}, I., {Pelkonen}, V.-M., {et~al.} 2011, \aap, 527,
  A111

\bibitem[{{Juvela} \& {Ysard}(2011)}]{Juvela2011}
{Juvela}, M. \& {Ysard}, N. 2011, \apj, 739, 63

\bibitem[{{Kawachi} \& {Hanawa}(1998)}]{Kawachi1998}
{Kawachi}, T. \& {Hanawa}, T. 1998, \pasj, 50, 577

\bibitem[{{K{\"o}hler} {et~al.}(2011){K{\"o}hler}, {Guillet}, \&
  {Jones}}]{Koehler2011}
{K{\"o}hler}, M., {Guillet}, V., \& {Jones}, A. 2011, \aap, 528, A96

\bibitem[{{K{\"o}hler} {et~al.}(2012){K{\"o}hler}, {Stepnik}, {Jones},
  {Guillet}, {Abergel}, {Ristorcelli}, \& {Bernard}}]{Koehler2012}
{K{\"o}hler}, M., {Stepnik}, B., {Jones}, A.~P., {et~al.} 2012, \aap, 548, A61

\bibitem[{{Kramer} {et~al.}(2003){Kramer}, {Richer}, {Mookerjea}, {Alves}, \&
  {Lada}}]{Kramer2003}
{Kramer}, C., {Richer}, J., {Mookerjea}, B., {Alves}, J., \& {Lada}, C. 2003,
  \aap, 399, 1073

\bibitem[{{Laureijs} {et~al.}(1991){Laureijs}, {Clark}, \&
  {Prusti}}]{Laureijs1991}
{Laureijs}, R.~J., {Clark}, F.~O., \& {Prusti}, T. 1991, \apj, 372, 185

\bibitem[{{Li} \& {Draine}(2001)}]{Li2001}
{Li}, A. \& {Draine}, B.~T. 2001, \apj, 554, 778

\bibitem[{{Lombardi} \& {Alves}(2001)}]{Lombardi2001}
{Lombardi}, M. \& {Alves}, J. 2001, \aap, 377, 1023

\bibitem[{{Martin} {et~al.}(2012){Martin}, {Roy}, {Bontemps},
  {Miville-Desch{\^e}nes}, {Ade}, {Bock}, {Chapin}, {Devlin}, {Dicker},
  {Griffin}, {Gundersen}, {Halpern}, {Hargrave}, {Hughes}, {Klein}, {Marsden},
  {Mauskopf}, {Netterfield}, {Olmi}, {Patanchon}, {Rex}, {Scott}, {Semisch},
  {Truch}, {Tucker}, {Tucker}, {Viero}, \& {Wiebe}}]{Martin2012}
{Martin}, P.~G., {Roy}, A., {Bontemps}, S., {et~al.} 2012, \apj, 751, 28

\bibitem[{{Men'shchikov} {et~al.}(2010){Men'shchikov}, {Andr{\'e}}, {Didelon},
  {K{\"o}nyves}, {Schneider}, {Motte}, {Bontemps}, {Arzoumanian}, {Attard},
  {Abergel}, {Baluteau}, {Bernard}, {Cambr{\'e}sy}, {Cox}, {di Francesco}, {di
  Giorgio}, {Griffin}, {Hargrave}, {Huang}, {Kirk}, {Li}, {Martin}, {Minier},
  {Miville-Desch{\^e}nes}, {Molinari}, {Olofsson}, {Pezzuto}, {Roussel},
  {Russeil}, {Saraceno}, {Sauvage}, {Sibthorpe}, {Spinoglio}, {Testi},
  {Ward-Thompson}, {White}, {Wilson}, {Woodcraft}, \&
  {Zavagno}}]{Menshchikov2010}
{Men'shchikov}, A., {Andr{\'e}}, P., {Didelon}, P., {et~al.} 2010, \aap, 518,
  L103

\bibitem[{{Nguyen Luong} {et~al.}(2011){Nguyen Luong}, {Motte}, {Hennemann},
  {Hill}, {Rygl}, {Schneider}, {Bontemps}, {Men'shchikov}, {Andr{\'e}},
  {Peretto}, {Anderson}, {Arzoumanian}, {Deharveng}, {Didelon}, {Di Francesco},
  {Griffin}, {Kirk}, {Konyves}, {Martin}, {Maury}, {Minier}, {Molinari},
  {Pestalozzi}, {Pezzuto}, {Reid}, {Roussel}, {Schuller}, {Testi},
  {Ward-Thompson}, {White}, \& {Zavagno}}]{Nguyen2011}
{Nguyen Luong}, Q., {Motte}, F., {Hennemann}, M., {et~al.} 2011, ArXiv e-prints

\bibitem[{{Nielbock} {et~al.}(2012){Nielbock}, {Launhardt}, {Steinacker},
  {Stutz}, {Balog}, {Beuther}, {Bouwman}, {Henning}, {Hily-Blant},
  {Kainulainen}, {Krause}, {Linz}, {Lippok}, {Ragan}, {Risacher}, \&
  {Schmiedeke}}]{Nielbock2012}
{Nielbock}, M., {Launhardt}, R., {Steinacker}, J., {et~al.} 2012, \aap, 547,
  A11

\bibitem[{{Ormel} \& {Cuzzi}(2007)}]{Ormel2007}
{Ormel}, C.~W. \& {Cuzzi}, J.~N. 2007, \aap, 466, 413

\bibitem[{{Ormel} {et~al.}(2011){Ormel}, {Min}, {Tielens}, {Dominik}, \&
  {Paszun}}]{Ormel2011}
{Ormel}, C.~W., {Min}, M., {Tielens}, A.~G.~G.~M., {Dominik}, C., \& {Paszun},
  D. 2011, \aap, 532, A43

\bibitem[{{Ormel} {et~al.}(2009){Ormel}, {Paszun}, {Dominik}, \&
  {Tielens}}]{Ormel2009}
{Ormel}, C.~W., {Paszun}, D., {Dominik}, C., \& {Tielens}, A.~G.~G.~M. 2009,
  \aap, 502, 845

\bibitem[{{Ossenkopf}(1993)}]{Ossenkopf1993}
{Ossenkopf}, V. 1993, \aap, 280, 617

\bibitem[{{Ossenkopf} \& {Henning}(1994)}]{Ossenkopf1994}
{Ossenkopf}, V. \& {Henning}, T. 1994, \aap, 291, 943

\bibitem[{{Ostriker}(1964)}]{Ostriker1964}
{Ostriker}, J. 1964, \apj, 140, 1056

\bibitem[{{Pagani} {et~al.}(2010{\natexlab{a}}){Pagani}, {Ristorcelli},
  {Boudet}, {Giard}, {Abergel}, \& {Bernard}}]{Pagani2010}
{Pagani}, L., {Ristorcelli}, I., {Boudet}, N., {et~al.} 2010{\natexlab{a}},
  \aap, 512, A3

\bibitem[{{Pagani} {et~al.}(2010{\natexlab{b}}){Pagani}, {Steinacker},
  {Bacmann}, {Stutz}, \& {Henning}}]{Pagani_core2010}
{Pagani}, L., {Steinacker}, J., {Bacmann}, A., {Stutz}, A., \& {Henning}, T.
  2010{\natexlab{b}}, Science, 329, 1622

\bibitem[{{Palmeirim} {et~al.}(2013){Palmeirim}, {Andr{\'e}}, {Kirk},
  {Ward-Thompson}, {Arzoumanian}, {K{\"o}nyves}, {Didelon}, {Schneider},
  {Benedettini}, {Bontemps}, {Di Francesco}, {Elia}, {Griffin}, {Hennemann},
  {Hill}, {Martin}, {Men'shchikov}, {Molinari}, {Motte}, {Nguyen Luong},
  {Nutter}, {Peretto}, {Pezzuto}, {Roy}, {Rygl}, {Spinoglio}, \&
  {White}}]{Palmeirim2013}
{Palmeirim}, P., {Andr{\'e}}, P., {Kirk}, J., {et~al.} 2013, \aap, 550, A38

\bibitem[{{Patanchon} {et~al.}(2008){Patanchon}, {Ade}, {Bock}, {Chapin},
  {Devlin}, {Dicker}, {Griffin}, {Gundersen}, {Halpern}, {Hargrave}, {Hughes},
  {Klein}, {Marsden}, {Martin}, {Mauskopf}, {Netterfield}, {Olmi}, {Pascale},
  {Rex}, {Scott}, {Semisch}, {Truch}, {Tucker}, {Tucker}, {Viero}, \&
  {Wiebe}}]{Patanchon2008}
{Patanchon}, G., {Ade}, P.~A.~R., {Bock}, J.~J., {et~al.} 2008, \apj, 681, 708

\bibitem[{{Pilbratt} {et~al.}(2010){Pilbratt}, {Riedinger}, {Passvogel},
  {Crone}, {Doyle}, {Gageur}, {Heras}, {Jewell}, {Metcalfe}, {Ott}, \&
  {Schmidt}}]{Pilbratt2010}
{Pilbratt}, G.~L., {Riedinger}, J.~R., {Passvogel}, T., {et~al.} 2010, \aap,
  518, L1

\bibitem[{{Planck Collaboration} {et~al.}(2011{\natexlab{a}}){Planck
  Collaboration}, {Abergel}, {Ade}, {Aghanim}, {Arnaud}, {Ashdown}, {Aumont},
  {Baccigalupi}, {Balbi}, {Banday}, {Barreiro}, {Bartlett}, {Battaner},
  {Benabed}, {Beno{\^i}t}, {Bernard}, {Bersanelli}, {Bhatia}, {Bock},
  {Bonaldi}, {Bond}, {Borrill}, {Bouchet}, {Boulanger}, \&
  {Bucher}}]{PlanckAbergel2011}
{Planck Collaboration}, {Abergel}, A., {Ade}, P.~A.~R., {et~al.}
  2011{\natexlab{a}}, \aap, 536, A25

\bibitem[{{Planck Collaboration} {et~al.}(2011{\natexlab{b}}){Planck
  Collaboration}, {Abergel}, {Ade}, {Aghanim}, {Arnaud}, {Ashdown}, {Aumont},
  {Baccigalupi}, {Balbi}, {Banday}, \& et~al.}]{PlanckMAMD2011}
{Planck Collaboration}, {Abergel}, A., {Ade}, P.~A.~R., {et~al.}
  2011{\natexlab{b}}, \aap, 536, A24

\bibitem[{{Planck Collaboration} {et~al.}(2011{\natexlab{c}}){Planck
  Collaboration}, {Ade}, {Aghanim}, {Arnaud}, {Ashdown}, {Aumont},
  {Baccigalupi}, {Balbi}, {Banday}, {Barreiro}, \& et~al.}]{PlanckBernard2011}
{Planck Collaboration}, {Ade}, P.~A.~R., {Aghanim}, N., {et~al.}
  2011{\natexlab{c}}, \aap, 536, A19

\bibitem[{{Planck Collaboration} {et~al.}(2011{\natexlab{d}}){Planck
  Collaboration}, {Ade}, {Aghanim}, {Arnaud}, {Ashdown}, {Aumont},
  {Baccigalupi}, {Balbi}, {Banday}, {Barreiro}, \& et~al.}]{PlanckMontier2011}
{Planck Collaboration}, {Ade}, P.~A.~R., {Aghanim}, N., {et~al.}
  2011{\natexlab{d}}, \aap, 536, A23

\bibitem[{{Poglitsch} {et~al.}(2010){Poglitsch}, {Waelkens}, {Geis},
  {Feuchtgruber}, {Vandenbussche}, {Rodriguez}, {Krause}, {Renotte}, {van
  Hoof}, {Saraceno}, {Cepa}, {Kerschbaum}, {Agn{\`e}se}, {Ali}, {Altieri},
  {Andreani}, {Augueres}, {Balog}, {Barl}, {Bauer}, {Belbachir}, {Benedettini},
  {Billot}, {Boulade}, \& {Bischof}}]{Poglitsch2010}
{Poglitsch}, A., {Waelkens}, C., {Geis}, N., {et~al.} 2010, \aap, 518, L2

\bibitem[{{Roberts} {et~al.}(2007){Roberts}, {Rawlings}, {Viti}, \&
  {Williams}}]{Roberts2007}
{Roberts}, J.~F., {Rawlings}, J.~M.~C., {Viti}, S., \& {Williams}, D.~A. 2007,
  \mnras, 382, 733

\bibitem[{{Roy} {et~al.}(2013){Roy}, {Martin}, {Polychroni}, {Bontemps},
  {Abergel}, {Andr{\'e}}, {Arzoumanian}, {Di Francesco}, {Hill}, {Konyves},
  {Nguyen-Luong}, {Pezzuto}, {Schneider}, {Testi}, \& {White}}]{Roy2013}
{Roy}, A., {Martin}, P.~G., {Polychroni}, D., {et~al.} 2013, \apj, 763, 55

\bibitem[{Skrutskie {et~al.}(2006)Skrutskie, Cutri, Stiening, Weinberg,
  Schneider, Carpenter, Beichman, Capps, Chester, Elias, Huchra, Liebert,
  Lonsdale, Monet, Price, Seitzer, Jarrett, Kirkpatrick, Gizis, Howard, Evans,
  Fowler, Fullmer, Hurt, Light, Kopan, Marsh, McCallon, Tam, Dyk, \&
  Wheelock}]{Skrutskie2006}
Skrutskie, M., Cutri, R., Stiening, R., {et~al.} 2006, aj, 131, 1163

\bibitem[{{Steinacker} {et~al.}(2010){Steinacker}, {Pagani}, {Bacmann}, \&
  {Guieu}}]{Steinacker2010}
{Steinacker}, J., {Pagani}, L., {Bacmann}, A., \& {Guieu}, S. 2010, \aap, 511,
  A9

\bibitem[{{Stepnik} {et~al.}(2003){Stepnik}, {Abergel}, {Bernard}, {Boulanger},
  {Cambr{\'e}sy}, {Giard}, {Jones}, {Lagache}, {Lamarre}, {Meny}, {Pajot}, {Le
  Peintre}, {Ristorcelli}, {Serra}, \& {Torre}}]{Stepnik2003}
{Stepnik}, B., {Abergel}, A., {Bernard}, J.-P., {et~al.} 2003, \aap, 398, 551

\bibitem[{{Stognienko} {et~al.}(1995){Stognienko}, {Henning}, \&
  {Ossenkopf}}]{Stognienko1995}
{Stognienko}, R., {Henning}, T., \& {Ossenkopf}, V. 1995, \aap, 296, 797

\bibitem[{{Stutz} {et~al.}(2010){Stutz}, {Launhardt}, {Linz}, {Krause},
  {Henning}, {Kainulainen}, {Nielbock}, {Steinacker}, \&
  {Andr{\'e}}}]{Stutz2010}
{Stutz}, A., {Launhardt}, R., {Linz}, H., {et~al.} 2010, \aap, 518, L87

\bibitem[{{van Breemen} {et~al.}(2011){van Breemen}, {Min}, {Chiar}, {Waters},
  {Kemper}, {Boogert}, {Cami}, {Decin}, {Knez}, {Sloan}, \&
  {Tielens}}]{vanBreemen2011}
{van Breemen}, J.~M., {Min}, M., {Chiar}, J.~E., {et~al.} 2011, \aap, 526, A152

\bibitem[{{Ward-Thompson} {et~al.}(1994){Ward-Thompson}, {Scott}, {Hills}, \&
  {Andre}}]{Ward-Thompson1994}
{Ward-Thompson}, D., {Scott}, P.~F., {Hills}, R.~E., \& {Andre}, P. 1994,
  \mnras, 268, 276

\bibitem[{{Weingartner} \& {Draine}(2001)}]{Weingartner2001}
{Weingartner}, J.~C. \& {Draine}, B.~T. 2001, \apj, 548, 296

\bibitem[{{Whittet} {et~al.}(1988){Whittet}, {Bode}, {Longmore}, {Adamson},
  {McFadzean}, {Aitken}, \& {Roche}}]{Whittet1988}
{Whittet}, D.~C.~B., {Bode}, M.~F., {Longmore}, A.~J., {et~al.} 1988, \mnras,
  233, 321

\bibitem[{{Wright}(1987)}]{Wright1987}
{Wright}, E.~L. 1987, \apj, 320, 818

\bibitem[{{Ysard} {et~al.}(2012){Ysard}, {Juvela}, {Demyk}, {Guillet},
  {Abergel}, {Bernard}, {Malinen}, {M{\'e}ny}, {Montier}, {Paradis},
  {Ristorcelli}, \& {Verstraete}}]{Ysard2012}
{Ysard}, N., {Juvela}, M., {Demyk}, K., {et~al.} 2012, ArXiv 1202.5966

\bibitem[{{Zhilkin} {et~al.}(2009){Zhilkin}, {Pavlyuchenkov}, \&
  {Zamozdra}}]{Zhilkin2009}
{Zhilkin}, A.~G., {Pavlyuchenkov}, Y.~N., \& {Zamozdra}, S.~N. 2009, Astronomy
  Reports, 53, 590

\bibitem[{{Zubko} {et~al.}(1996){Zubko}, {Mennella}, {Colangeli}, \&
  {Bussoletti}}]{Zubko1996}
{Zubko}, V.~G., {Mennella}, V., {Colangeli}, L., \& {Bussoletti}, E. 1996,
  \mnras, 282, 1321

\end{thebibliography}

\appendix

\section{Estimated bias of extinction maps}

The resolution of the NICER extinction maps is 200$\arcsec$. This is not much smaller than the width of the filaments and can bias the estimates of peak extinction because more background stars are visible
through the filament edges where the column density is lower. To estimate the magnitude of the error, we analysed a set of simulated observations. We started with the observed $A_{\rm V}$ profile of the
first cut and used that to construct a two-dimensional image of a filament with the same extinction profile. We took 2MASS stars observed towards a nearby low-extinction field and distributed them
over the image with a stellar density that is similar to the actual Taurus observations. The signal of the background stars was attenuated and the photometric errors were increased to correspond to the noise
levels and the detection thresholds of the Taurus field. The modified magnitudes were fed to the NICER routine to construct an ``observed'' extinction map.

\begin{figure}[!ht]
\centerline{
\includegraphics[width=0.3\textwidth]{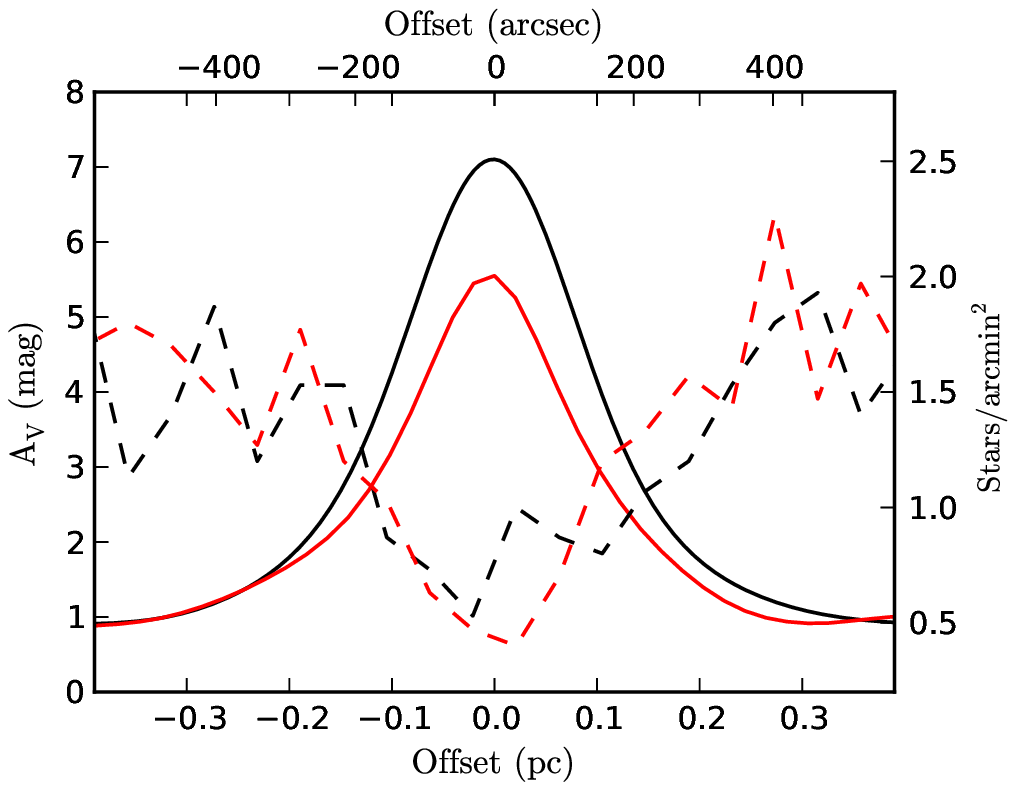}}
\caption{Simulation of possible bias in NICER estimates. The black solid curve shows the assumed true $A_{{\rm V}}$ profile, convolved to a resolution of 200$\arcsec$. The red solid line shows the extinction profile recovered with NICER algorithm from the simulated observations. The dashed lines and the right hand scale indicate the stellar density in the case of the first cut (black dashed line) and in the simulation (red dashed line).}
\label{appendix_simu} 
\end{figure}

Figure~\ref{appendix_simu} shows as a solid black line the input $A_{{\rm V}}$ profile convolved to the resolution of 200$\arcsec$, which has a maximum of $\sim 7^{\rm m}$. The red line shows the $A_{{\rm V}}$ profile recovered from the calculated NICER map, which has a maximum of $\sim 5.5^{\rm m}$. The noise is low because we have averaged data over wider strips across the simulated filament. The peak extinction is underestimated by $\sim 21$\%, while the contrast between the filament and the assumed background extinction is lower by $\sim 25$\%. The recovered peak extinction of $\sim 5.5^{\rm m}$ magnitudes, for an angular resolution of 200\arcsec, is significantly lower than the assumed true peak value of $\sim 9^{\rm m}$ magnitudes at full angular resolution. This is in roughly equal parts due to the bias and due to the natural effect of a change in the spatial resolution. This also means that the simulation is only a conservative estimate of the true bias. If we repeat this simulation for a filament with a lower visual extinction ($A_{\rm V} = 5^{\rm m}$ at 200\arcsec angular resolution), the recovered peak extinction is still underestimated by 16\%.

\end{document}